\documentstyle[preprint,aps]{revtex}

\begin{document}
\draft
\title{Bound states and impurity averaging
in unconventional superconductors}
\author{Robert Joynt}
\address{Department of Physics \\
University of Wisconsin-Madison \\
1150 University Avenue \\
Madison, WI 53706 \\}
\date{\today}
\maketitle

\begin{abstract}
The question of anomalous transport due to 
a band of impurity states in unconventional 
superconductors is discussed.  In general, the 
bound state energies are not in midgap, even in the
unitarity limit.  This implies that, generically, the
states associated with impurities are broad resonances,
not true bound states.  There is no impurity band in
the usual sense of the phrase.
The wavefunctions of these resonances
possess interesting anisotropies in real space, but this 
does not result in anomalous hopping between impurities.
I conclude that the system of resonances produces no
qualitative modifications to the T-matrix theory 
with impurity averaging which is
normally used to treat the low-temperature transport
of unconventional superconductors.
However, users of this method often assume a density of states
which is symmetric around the chemical potential.  This is not
normally the case.  It is found that 
the non-crossing approximation is 
not valid in a strictly two-dimensional system.
\end{abstract}
\section{Introduction}
Impurity scattering plays a dominant role in the transport
and thermodynamic properties of unconventional superconductors,
far larger than in conventional s-wave superconductors.
This is a consequence of the gap nodes which prevent the complete freezing
out of scattering processes, and the fact that an anisotropic order
parameter is far more sensitive to disorder.  The critical temperature for an 
anisotropic superconductor is suppressed even in lowest order by
the disorder potential.  This follows from the breakdown of one of the
conditions for Anderson's theorem \cite{anderson}, which is that the momentum 
dependence of the pair potential is weak.  A breakdown of the theorem 
leads to bound states in the gap when there are magnetic impurities in
s-wave materials.  It also leads to the possibility
of such bound states from nonmagnetic impurities in the gap of an
unconventional superconductor.  This paper is devoted to
questions about these states: their energy levels,
their wavefunctions, their lifetimes,
and the role (if any) they play in observable properties at low temperatures.

This subject is topical because of the interest in high-temperature
superconductors.  Some of these systems appear to have gap nodes,
implying the presence of an unconventional order parameter.
No picture of these materials is complete without understanding the effects of 
dirt.  Furthermore, experiments in the asymptotic low-temperature regime
are special in that they probe the region of the Fermi surface near the 
nodes.  It is in this region where the effects of impurities are most dramatic.
The same considerations hold for the comparatively venerable 
heavy-fermion superconductors.  Here we have solid grounds for supposing that 
some of these systems, particularly UPt$_3$ and UBe$_{13}$, are unconventional.
Still, after more than a decade of investigation, the experimental details of the
thermodynamic and transport properties of these systems
at low temperatures are not fully
reconciled with theory.  In heavy fermion materials, 
however, it has become clear that
strong impurity scattering, approaching the unitary limit, is the
rule, not the exception.  The Born approximation is inadequate.
In high-T$_c$ systems, this is still under debate.
In this paper, I will concentrate on this near-unitary limit.

The literature on impurity states in the superconducting gap begins with
the papers of Yu \cite{yu} and Shiba \cite{shiba} 
on magnetic impurities in s-wave
systems.  The two important ingredients are the pairbreaking nature of the 
disorder potential and the 'hard' energy gap - the density of states (DOS)
is zero in some neighborhood of the chemical potential in the pure system.
Bound states appear in the gap.  Increasing the impurity concentration
increases the number of bound states and decreases the gap, leading first to
the gapless state and finally to the destruction of superconductivity
\cite{maki}.

In unconventional superconductors, the bound states arising from ordinary 
potential scattering were first considered by Buchholtz and Zwicknagl 
\cite{buch}.  They concentrated on the Balian-Werthamer state, which
has a hard energy gap, but the momentum-averaged gap vanishes:
$\sum_{\vec{k}} \Delta(\vec{k}) = 0$, where the sum is over the Fermi surface.
For such a gap, the results are somewhat similar to the previous case as the 
disorder potential is likewise pairbreaking.  Although time-reversal symmetry is 
not broken, the randomization of momentum in the eigenstates prevents pairing 
by a momentum-dependent potential.  Bound states appear in the gap.  
These authors also state that, in the unitarity limit of very strong 
potentials on the impurities, the bound states are at midgap.  This statement 
has been repeated many times in the literature.  However, I will argue below 
that it is incorrect.

One important point about unconventional superconductors is that they cannot 
exist at very high impurity density.  The critical temperature decreases as the 
impurity concentration is increased, and vanishes when $\hbar/\tau \approx k_B 
T_{c0}$ where $T_{c0}$ is the critical temperature in the absence of scattering. 
This implies that the regime of low impurity density is the only one of 
interest.

With the discovery in the 1980's of the heavy fermion superconductors, there 
was an explosion of interest in the problem of disorder in unconventional 
superconductors.  Many calculations of transport and thermodynamic 
properties at low temperatures have been published \cite{sigrist}.
The standard method, explained most completely by Hirschfeld {\it et al.} 
\cite{hirschfeld}, combines the T-matrix approximation with standard impurity 
averaging techniques.  Generally speaking, scattering near or at the unitarity 
limit is required to explain experiments in both the heavy-fermion 
\cite{sigrist}, \cite{heat1} and perhaps also in the 
high-T$_c$ materials \cite{heat2}, \cite{heat3}.  
This suggestion that the unitary limit is the appropriate one
for Kondo lattice systems is due to Pethick and Pines \cite{pethick}.
While this model is certainly relevant to the weakly hybridizing
$f$-level electrons in heavy fermion materials, its applicability to high-T$_c$ 
systems is unclear.

The superconducting order parameters considered for 
both kinds of systems do satisfy
the $\sum_{\vec{k}} \Delta(\vec{k}) = 0$ condition, but they do not have a 
hard energy gap.  The DOS of the pure system is usually taken to vanish 
linearly or quadratically at the chemical potential $\mu$.  The standard method 
of treating the disorder potential leads to a finite DOS at $\mu$ \cite{ueda}.
The neighborhood of the chemical potential where the density of states is
flat is sometimes referred to as the 'impurity band' \cite{heat3}.

The impurity averaging method for unconventional superconductors has been
explicitly questioned by some recent 
work \cite{balatsky}.  The gap nodes lead to unusual 
wavefunctions for the bound states, with the possibility of anomalous overlaps 
between well-separated impurities.  In compensated doped semiconductors, a high 
concentration of impurities can lead to a new conduction mechanism which
predominates at low 
temperatures, conduction entirely through the impurity wavefunctions which 
form the impurity band \cite{mott}.  This possibility must be considered also in 
superconductors.  The electrical conductivity of any such band would of course
be shorted out by the conductivity of the condensate, but the opposite could 
well occur for the thermal conductivity.  I will argue below that this does 
{\it not} occur.

A more radical criticism of impurity averaging
for two-dimensional sysyems 
is contained in papers of Neresyan {\it et al.} \cite{neresyan}, who find that 
multisite processes restore the vanishing of the DOS at the chemical 
potential.  A recent preprint of Ziegler {\it et al.} 
\cite{ziegler} shows that, for Lorentzian 
disorder, the finite DOS is not a consequence of impurity averaging.  

While the aim of the current work is to clarify the theoretical situation for
nonmagnetic impurities in unconventional superconductors, there 
has been considerable recent
work on magnetic impurities in both conventional and unconventional
superconductors, stimulated by experiments \cite{yazdani}. 
Some of this work has reached conclusions similar to those 
presented here, particularly with regard to the importance of carefully
considering the real part of the Green's function in T-matrix
calculations \cite{salkola}, \cite{flatte}.  

In order to build up the theory from the start, I begin in Sec.\ \ref{sec:semi}
with the question of bound states in the normal state of a semiconductor with a 
gap, in the limit of strong scattering.  Since the semiconductor analogy is 
a powerful (but not omnipotent) one, this section provides much of the 
basis for the paper.  
The s-wave case is treated briefly in Sec.\ \ref{sec:swave}, 
both to establish notation and to get a basis of comparison
with unconventional superconductivity.  
This latter topic, the main subject of the paper,
is begun with calculations of the wavefunctions and lifetimes for single
impurities in d-wave-type systems in Sec.\ \ref{sec:dwave}.  
Finally, in Sec.\ \ref{sec:many}
the many-impurity case is discussed, along with the experimental
implications for real systems.   

\section{Bound states in semiconductors}
\label{sec:semi}
\subsection{Introduction and formalism}

I examine an imaginary semiconductor in this section.
The goal is to understand the process of binding an electron
to an impurity with a very strong short-range potential.
The physics of this process is sufficiently different from the
textbook cases that certain features are likely to
be as unfamiliar to the reader as they were to the writer.
These features are important for the superconducting
model which is believed to be
of most relevance for high-T$_c$ and heavy fermion
superconductors.

Let us consider a semiconductor with a single impurity.
The gap is the result of the lattice potential, a 
single-particle effect, 
and is not tied to the chemical potential, which lies in the gap.  
In the limit of weak scattering, this is essentially the familiar
case of off-valence impurities in a Group IV material.  This leads to impurity 
states very near the band edges.
Our interest is in the opposite limit when the scattering is strong.
The unperturbed Hamiltonian is 
\begin{equation}
\hat{H}_0 = \sum_{\vec{k}} \epsilon_{\vec{k}} n_{\vec{k}}.
\end{equation}
The energies are measured from the chemical potential.
I have omitted band and spin
indices for clarity.  The sum over momentum 
is always taken to 
include a sum over bands.
The potential for a single short-range impurity is
\begin{equation}
\hat{V} = V \sum_{\vec{k},\vec{k}'} c^{\dagger}_{\vec{k}}c_{\vec{k}'}.
\end{equation}
This is an s-wave potential.  The phase shift is
\begin{equation}
\delta(\epsilon) = - \tan^{-1}(\pi  N_0(\epsilon) V),
\end{equation}
where $N_0(\epsilon)$ is the density of states
of the unperturbed system. 
The phase shift normally quoted in papers on
transport properties is for states 
at
the Fermi surface:
\begin{equation}
\delta_0 = - \tan^{-1}[\pi  N_0(\epsilon_F) V],
\end{equation}
The unitarity limit 
$\delta \rightarrow \pi/2 $ is reached when
$V \rightarrow - \infty$.

The unperturbed Green's function is
\begin{equation}
{\cal G}_0(\vec{k},i \omega) = \frac{1}{i \omega - \epsilon_{\vec{k}}}         
\end{equation}
The equation for the full Green's function for the Hamiltonian 
$\hat{H} = \hat{H}_0 + \hat{V}$ is
\begin{eqnarray}
{\cal G}(\vec{k},\vec{k}', i \omega) & = & {\cal G}_0(\vec{k}, i \omega) \delta_{\vec{k},\vec{k}'}
                          + {\cal G}_0(\vec{k}, i \omega) V {\cal G}_0(\vec{k}', i \omega)
                           + {\cal G}_0(\vec{k}, i \omega) V \sum_{\vec{k}_1}
                                 {\cal G}_0(\vec{k}_1, i \omega) 
                                V {\cal G}_0(\vec{k}', i \omega) 
                          + \ldots  \nonumber         \\
                          & = & {\cal G}_0(\vec{k}, i \omega) 
                                \delta_{\vec{k},\vec{k}'}
                                + {\cal G}_0(\vec{k}, i \omega) T(\omega) 
                                   {\cal G}_0(\vec{k}', i \omega).                                    
\label{eq:gfull}
\end{eqnarray}                                                                        
Here
\begin{eqnarray}
T(i \omega) & = & V + V^2 \sum_{\vec{k}} {\cal G}_0(\vec{k}, i \omega)
                +V^3 [\sum_{\vec{k}} {\cal G}_0(\vec{k}, i \omega)]^2 + \ldots \\
          & = & \frac{V}{ 1 - V g_0(i \omega)}       
\end{eqnarray}
and
\begin{equation}                                                                            
g_0(i \omega) \equiv \sum_{\vec{k}} \frac{1}{i \omega - \epsilon_{\vec{k}}}.
\end{equation}
Continuing this function to the real axis:
\begin{equation}
g_0(\omega+ i \delta) = \sum_{\vec{k}} \frac{1}{ \omega - \epsilon_{\vec{k}} + 
i \delta} = \sum_{\vec{k}} \frac{{\cal P}}{ \omega - \epsilon_{\vec{k}}}
- i \pi \sum_{\vec{k}} \delta(\omega - \epsilon_{\vec{k}}).
\end{equation}                                                                 
Thus 
\begin{equation}
Im~ g_0(\omega+ i \delta) 
= -  \pi \sum_{\vec{k}} \delta(\omega - \epsilon_{\vec{k}})
= - \pi N_0(\omega),
\end{equation}
where $N_0(\omega)$ is the density of states for one spin.
Also
\begin{equation}
Re~ g_0(\omega+ i \delta) = {\cal P} \int \frac{N_0(\omega')}{\omega - \omega'} 
d \omega',
\end{equation}
proportional to the Hilbert transform of the density of states.

The expression
\begin{equation}
T(\omega) = \frac{V}{ 1 - V g_0(\omega)}
\end{equation}
shows that $T$ has poles only when $g_0$ is purely real and 
\begin{equation}
\frac{1}{V} = Re~g_0(\omega_b).
\label{eq:pole}
\end{equation}
Let us agree that when real frequency arguments are used, a limit
is implied where the real frequency axis is approached from above
in the complex plane,
corresponding to retarded functions.
When Eq.\ \ref{eq:pole} is satisfied but $g_0$
has an imaginary part, then $T$ is a Lorentzian near $\omega_b$,
and we are dealing with a resonance.
If there is a pole, it represents a bound $(V < 0)$ or an
antibound $(V > 0)$ state.  In these cases we may write
\begin{equation}
T^{-1}(\omega \approx \omega_b) \approx \frac{1}{V} - g_0(\omega_b)
- (\omega - \omega_b) g_0'(\omega_b) = - (\omega - \omega_b) g_0'(\omega_b), 
\end{equation}
where
\begin{equation}
g_0'(\omega_b) = - {\cal P} \int \frac{N_0(\omega') d \omega'}
{(\omega_b - \omega')^2} \equiv - Z.
\label{eq:z}
\end{equation}
The integral for $Z$ always converges because
$N_0(\omega_b) = 0$ for a true bound state.
Therefore we find
\begin{equation}
T(\omega \approx \omega_b) 
\approx \frac{Z^{-1}}{\omega - \omega_b },
\end{equation}
in the neighborhood of the pole.
The bound states are therefore characterized by poles in the T-matrix,
and resonances by a sharp peak in the imaginary part of the T-matrix.

The T-matrix gives the exact solution for 
the one impurity problem.  It is not 
an approximation.

\subsection{Case of symmetric bands}
Let us consider a semiconductor with 
an unperturbed density of states which has 
a gap of width $2 \Delta$, 
and satisfies the symmetry relation
\begin{equation}
N_0(\omega) = N_0(- \omega).
\end{equation}
I will argue below that this is not 
likely to be realized in the cases of interest, but
it is the simplest mathematically.
The density of states is illustrated in Fig.\ \ref{fig:symm}.
Now we have
\begin{equation}
Re~g_0(\omega) = {\cal P} \int 
\frac{N_0(\omega') d \omega'}{\omega - \omega'} 
\approx - \frac{2 \omega}{\omega_g^2},
\end{equation}
in the region $|\omega| << \omega_g$,
where
\begin{equation}
\omega_g^{-2} \equiv \int  \frac{N_0(\omega') d \omega'}{\omega'^2} > 0.
\end{equation}
$\omega_g$ is of order $ \Delta$ if the bandwidth is
smaller than the gap energy (characteristic of insulators)
and is of order the geometric
mean of bandwidth
times the gap energy in the other limit where the bandwidth is
much greater than the gap energy (characteristic of semiconductors).  
The bound state energy 
$\omega_b$ satisfies
\begin{equation}
\frac{1}{V} =  - \frac{2 \omega_b}{\Delta^2},
\end{equation}
or
\begin{equation}
\omega_b = - \frac{2 \Delta^2}{V}.
\end{equation}
For large $|V|$, (the unitary limit) this is a midgap state.
This limit is shown in Fig.\ \ref{fig:symm}.
For $V<0$ the potential is attractive and the state sits just above the
middle of the gap.  This is an ordinary bound state.  If $V>0$, it 
sits just below the 
center of the band.
It is an 'antibound' state, but the wavefunction is localized, just as for
a bound state.
     
\subsection{Density of states}
The density of states is
\begin{equation}
N(\omega) = - \frac{1}{\pi} Im~Tr~{\cal G}
(\vec{k},\vec{k}', \omega + i \delta).
\end{equation}
Comparing this with the equation
\begin{equation}
{\cal G}(\vec{k},\vec{k}', \omega) =
{\cal G}_0(\vec{k}, \omega) \delta_{\vec{k},\vec{k}'}
+ {\cal G}_0(\vec{k}, \omega) T(\omega) {\cal G}_0(\vec{k}', \omega),
\end{equation}
we find in the gap region (when $Im~{\cal G}_0 = 0$):
\begin{equation}
N(\omega) = N_0(\omega) - \frac{1}{\pi} \sum_{\vec{k}} 
(\omega - \epsilon_{\vec{k}})^{-2} Im~T(\omega),
\end{equation}
which may be written in terms of the change in the density of states:
\begin{equation}
\Delta N(\omega) =  N(\omega) -  N_0(\omega)
= - \frac{1}{\pi} \sum_{\vec{k}} 
(\omega - \epsilon_{\vec{k}})^{-2} Im~T(\omega).
\end{equation}
Near $\omega = \omega_b$ this expression yields
\begin{equation}
N_{imp} = - \frac{1}{\pi} [\sum_{\vec{k}} (\omega - \epsilon_{\vec{k}})^{-2}]
Im~\frac{Z^{-1}}{\omega - \omega_b + i \delta} = \delta(\omega - \omega_b),
\end{equation}
which is the impurity contribution to the density of states.
To obtain the second equality, I have used
Eq.\ \ref{eq:z}.
In the region where $Im~g_0(\omega) \neq 0$, we also have that
$Im~T(\omega) \neq 0$, and this represents a phase shift with
an accompanying reduction of the density of states
of the continuum such that
\begin{equation}
\int \Delta N(\omega) d \omega = 0.
\end{equation}
The reduction of the density of states of the continuum 
just cancels the additional bound state (Levinson's theorem). 

\subsection{Local density of states}

Near the bound state energy,
the Green's function in real space has the form
\begin{equation}
\frac{|\psi(\vec{r})|^2}{\omega - \omega_b + i \delta}.
\end{equation}
Comparing with Eq.\ \ref{eq:gfull}, 
we may extract the wavefunction $\psi(\vec{r})$
by taking the Fourier transform of $G_0(\omega + i \delta)$:
\begin{equation}
\psi(\vec{r}) \sim 
\int \frac{e^{i \vec{k} \cdot \vec{r}} d^3 k}
{\omega_b - \epsilon_{\vec{k}} + i \delta}. 
\label{eq:psi}
\end{equation}
Consider a semiconductor with a bound state 
at $\omega_b$ and band edges at $ \pm \Delta/2$.
Let the bands be parabolic.
Then the contribution from the upper band is:
\begin{eqnarray}
\psi(\vec{r}) & \sim &  \int 
\frac{e^{i \vec{k} \cdot \vec{r}} d^3 k}
{\omega_b - (\Delta/2 + k^2/2m)} \\
  & = &   - 2m  \int 
\frac{e^{i \vec{k} \cdot \vec{r}} d^3 k}
{k^2 + k_0^2} \\
  & = &   -  4 \pi m  \int_0^{\infty} k^2 dk \int_{-1}^1 dx
\frac{e^{i k r x}} {k^2 + k_0^2} \\
  & = &   -\frac{4 \pi m}{i r} \int_0^{\infty} k dk 
\frac{e^{i k r } - e^{ - i k r}} {k^2 + k_0^2} \\
  & = &   -  \frac{4 \pi^2 m } {r} e^{- k_0 r},
\label{eq:ko}
\end{eqnarray}
where $k_0^2  \equiv m \Delta - 2 m \omega_b$.
The contribution from the lower band is the same except that
$k_0$ is replaced by $|m \Delta + 2 m \omega_b|$.
The wavefunction is very tightly bound,
the decay length being short because the energy
is far from the band edge.

In the limit of a very weak attractive potential
($V < 0$ and $V N_u <<1$), then
we are interested in the form of $Re~g_0(\omega)$ when
$\omega \approx \Delta$.  We find
\begin{equation}
Re~g_0(\omega) \approx N_u \log ( \frac{ \Delta - \omega}{\epsilon_u}).
\end{equation}
The bound state energy is:
\begin{equation}
\omega_b = \Delta - \epsilon_u e^{1/N_u V}.
\end{equation}
This is a state just below the upper band.  The lower band has no effect in this 
case.  It is important to note that the exponential dependence 
for the bound state energy is
due to the fact that there is a finite jump in the density of states
at the band edge.  If there is a square root singularity:
\begin{equation}
N_0(\omega) \sim (\omega - \Delta)^{1/2}
\end{equation}
for $\omega > \Delta$,
as one would expect in three dimensions, then there is a threshold
coupling strength below which there is no bound state.    

The hydrogenic impurity case, of great practical importance, 
is different from all of these cases 
because of the long-range potential, which 
leads to an infinite number of bound 
states for all interaction strengths even in three dimensions.

\subsection{Case of asymmetric bands}

If the bands are asymmetric, 
$N_0(\omega) \neq N_0(- \omega)$,
then {\it the bound state is not in the middle of the
gap even when} $|V| \rightarrow \infty$.
This result is illustrated in Fig.\ \ref{fig:asymm}.
Consider an example where the lower band extends from
$-\epsilon_{\ell}$ to $-\Delta$, and the upper 
band                               
from $\Delta$ to $\epsilon_u$.  Let the bands have constant density
of states $N_{\ell}$ and $N_u$, respectively.
Then                                 
\begin{eqnarray}
Re~g_0(\omega) & = & N_{\ell} \int_{-\epsilon_{\ell}}^{-\Delta} 
\frac{d \omega'}{\omega - \omega'} + 
N_{u} \int_{\Delta}^{\epsilon_u} 
\frac{d \omega'}{\omega - \omega'} \\
& = & - N_{\ell} \log | \frac{\omega + \Delta}{\epsilon_{\ell}}|
      + N_u \log | \frac{\omega - \Delta}{\epsilon_u}| .
\end{eqnarray}
If the enrgy is in the gap,
$|\omega| < \Delta$, then this may be written as
\begin{equation}
Re~g_0(\omega) = 
        N_{\ell} \log | \frac{\epsilon_{\ell}}{\Delta}|
      - N_u \log | \frac{ \epsilon_u}{\Delta}|
        - \frac{N_u + N_{\ell}}{\Delta} \omega.
\end{equation}
The bound state equation $ Re~g_0(\omega) = 1/V$ now has the  
solution:
\begin{equation}
\omega_b = - \frac{\Delta}{N_u + N_{\ell}}
\left[\frac{1}{V} + N_u \log(\frac{ \epsilon_u}{\Delta})
-N_{\ell} \log(\frac{ \epsilon_{\ell}}{\Delta}) \right].
\end{equation}
Even in the limit of very strong scattering, this is not a midgap state.
The bound state energy is 
displaced away from the band with the higher density of states
because of level repulsion.
The asymptotic behavior of the wavefunction
(the radius of the bound state)
is still determined by the distance to the nearest band edge and the 
effective mass of that band.  

The effect of the asymmetry may be described as 
a renormalization of the potential in the following way.  
We may rewrite the 
eigenvalue equation as:
\begin{equation}
\omega_b = - \frac{\Delta} {(N_u + N_{\ell})\tilde{V}},
\end{equation}
if we define the renormalized potential strength as
\begin{equation}
\tilde{V} = \frac{V}{1+ V N_u \log(\frac{ \epsilon_u}{\Delta})
-V N_{\ell} \log(\frac{ \epsilon_{\ell}}{\Delta})}
\equiv \frac{V}{1 + V N_A}.
\end{equation}
This equation defines the asymmetry factor $N_A$,
\begin{equation}
N_A =  N_u \log\left(\frac{ \epsilon_u}{\Delta}\right)
-N_{\ell} \log\left(\frac{ \epsilon_{\ell}}{\Delta}\right)
\end{equation}
which is of the same order of 
magnitude as the density of states 
at the Fermi energy.
It is $\tilde{V}$ not $V$, that determines the energy of the 
bound state.
It is important to note that, as $V \rightarrow \infty$,
\begin{equation}
\tilde{V} \rightarrow \frac{1}{N_A}.
\end{equation}
If the upper band is dominant, $N_A>0$, and the potential is repulsive,
$V>0$, then we have that $\tilde{V} < V$ and the antibound state always 
stays below the center of the gap.  This is simply a consequence
of level repulsion.  There is a similar effect for $V<0$ and $N_A <0$,
with the bound state never reaching the center of the gap even if
$V \rightarrow \infty$.  For the other combinations of signs,
we will have a midgap state only in the 'accidental' case
that $V = - 1/N_A$.

This issue of band symmetry is crucial
for the understanding of the bound state problem.
It is particularly important to distinguish
band symmetry from particle-hole symmetry,
which is a very useful approximation for many
calculations in superconductivity theory.
Particle-hole symmetry is the assumption,
approximately true in most cases, that the 
density of states of the normal material
does not vary appreciably in the neighborhood
of the Fermi energy, the neighborhood being here defined as
the range of energies within the cutoff energy $\hbar \omega_c$
for the pairing interaction.  The approximation may be stated as
\begin{equation}
\omega_c \frac{d N_0(\omega)}{d \omega}(\omega = \epsilon_F) << N_0(\epsilon_F)
\end{equation} 
This is used in many
elementary calculations of superconducting properties
because only this range of energies is important
for many purposes.  A good example is the calculation of the 
critical temperature in the weak-coupling theory. 
The validity of the approximation arises ultimately
from the mismatch
of electronic and phononic (or other bosonic) 
time scales.  

Band symmetry is the assumption  $N_0(\omega) = N_0(-\omega)$
{\it which is essentially never valid}.
To give an idea of how far it fails,
I have computed numerically the 
asymmetry factor for the following model semiconductor.
It is a two-dimensional square lattice
with a nearest-neighbor hopping 
matrix element $t$ and 
a filling of 0.8 electrons per unit
cell.  The dispersion is $\epsilon_{\vec{k}} =
- 2.0 t [\cos(k_x) + \cos(k_y)]$.
At the Fermi energy $\epsilon_F = -0.4 t$, there is a 
gap, symmetric around $\epsilon_F$,
of $0.02 t$.  The density of states is shown in 
Fig.\ \ref{fig:tbgap}.  Then $N_A$ is defined as
\begin{equation}
N_A = \langle \frac{1}{\epsilon_{\vec{k}} - \epsilon_F} \rangle, 
\end{equation}
where the brackets indicate an average over the band.
The result is $N_A = 0.25/t$ per unit cell.
Since the total band width is $W = 8 t$, we see that
the product $N_A W $ is of order unity.  
For any band, $N_A$ as a function of filling has one 
zero at
some point.  For this particular band, this occurs at half
filling.  In general, however, it is only for a 
special choice of $\epsilon_F$ that $N_A = 0$
and the band symmetry assumption is valid.  The physical
distinction between particle-hole and band symmetry 
is that there is no frequency mismatch for impurity scattering.
The ionic potential which produces the band 
structure and the impurity potential are instantaneous.
It is natural, but completely unjustifiable, to extend
particle-hole symmetry to band symmetry.  No conclusion which is
based on such an extension 
is likely to apply to any real material.

\subsection{Level occupation}

In a semiconductor, the occupation of impurity 
levels is normally strongly dependent on the 
valence of the impurity relative to the valence
of the constituent atoms.  Here, we have been using a 
model 
in which the valence of the impurity and the background
atoms is the same.  
We explicitly do not introduce additional states,
only a potential which moves the old states around.  
This distinguishes the present
work from Anderson magnetic impurity models, which generally do
introduce such new states.

The occupation at zero temperature
is then as follows.  If the potential is repulsive, the 
impurity 'peels off' one state from the valence band.
It is therefore full, regardless of its position in 
the gap, and even if it is above midgap.
If the potential is attractive, the 
impurity peels off one state from the valence band.
It is therefore empty, also regardless of its position in 
the gap.  In real semiconductors, it is normally true
that the Coulomb repulsion prevents double occupancy
of impurity levels, an effect not considered here.  

\subsection{Many impurities}
\label{sec:mase}
\subsubsection{Impurity band formation}

If there are $N_{imp}$ impurities
at a finite density $n_{imp}$
in the system, we
must consider the possibility that the wavefunctions 
on different impurities overlap.  
We begin with the case of
two impurities.
The Hamiltonian is then 
\begin{equation}
\hat{H} = \hat{H}_0 + \hat{V}_1 + \hat{V}_2.
\end{equation}
$\hat{H}_0 $ is the Hamiltonian of the pure system,
$\hat{V}_1$ is the potential of the impurity at
site $\vec{r}$, and $\hat{V}_2$ is the potential of the 
impurity at the site $\vec{r}+\vec{R}$.  Our interest is in the limit
$R k_0 >>1$, where $k_0$ is the inverse of the bound state radius, as in
Eq.\ \ref{eq:ko}.
The bound state wavefunctions satisfy
\begin{equation}
(\hat{H}_0 + \hat{V}_1)\psi_1 = \omega_b \psi_1
\end{equation}
and
\begin{equation}
(\hat{H}_0 + \hat{V}_2)\psi_2 = \omega_b \psi_2.
\end{equation}
The overlap matrix element is
\begin{equation}
M_{12} = \langle \psi_2 | \hat{H} | \psi_1 \rangle = 
         \omega_b \langle \psi_2  | \psi_1 \rangle + 
         \langle \psi_2 | \hat{H}_1 | \psi_1 \rangle.
\end{equation}
The two terms are generally of the same size and asymptotic
behavior.  Taking the first as representative, we find
\begin{equation}
M_{12} \sim \omega_b \langle \psi_2 | \psi_1 \rangle \sim
\omega_b \int \psi^*(\vec{r}) \psi(\vec{r}+\vec{R}) d^3r,
\end{equation}
where $\psi$ is the impurity wavefunction.
Using Eq.\ \ref{eq:psi}, we find, for the symmetric case
\begin{eqnarray}
M_{12}(\vec{R}) 
& \sim & 
\omega_b \int \left[\frac{e^{i \vec{k} \cdot \vec{r}} d^3 k}
{\omega_b - \epsilon_{\vec{k}}}
\int \frac{e^{i \vec{k}' \cdot (\vec{r}+\vec{R})} d^3 k'}
{\omega_b - \epsilon_{\vec{k}'}} \right] d^3 r \nonumber \\ 
& \sim &
\omega_b
\int \frac{e^{i \vec{k} \cdot \vec{R}} d^3 k}
{\left[\omega_b - \epsilon_{\vec{k}} \right]^2} \nonumber \\ 
& = & \frac{8 \pi^2 m \omega_b}{k_0} e^{-k_0R}. 
\end{eqnarray}
As in Eq.\ \ref{eq:ko}, we have the decay length
$k_0^2  \equiv m \Delta - 2 m \omega_b$.
We may now write a Hamiltonian for the many-impurity case
in the basis of the bound state wavefunctions at different
sites.  The resulting impurity bandwidth is 
of order $\omega_b \exp(-k_0 n_{imp}^{-1/3})$.
$\omega_b$ is less than the gap energy $\Delta$
and $1/k_0$ is of the order of the lattice spacing.
We generally expect a {\it very} small bandwidth
for this 'deep impurity' ($\omega_b \sim \Delta$)
case.  There is therefore no metallic conduction
when there is even a small amount of disorder 
in the impurity site energies.  Interactions will also 
tend to localize the electrons and strengthen this conclusion.

\subsubsection{Impurity averaging}

The Hamiltonian for the many-impurity case is
\begin{equation}
\hat{H} = \hat{H}_0 + \sum_i V(\vec{r} - \vec{R}_i).
\label{eq:himp}
\end{equation}
The impurities are located at the position $\vec{R}_i$.
The standard method of calculation is to average over the 
positions $\vec{R}_i$ (impurity averaging) \cite{agd}.
This method is valid for calculating the effects 
of impurities on the existing 
states if there are no correlations in the quantities 
$\psi_0^*(\vec{R}_i)\psi_0(\vec{R}_j)$, where $\psi_0^*(\vec{r})$
are the eigenfunctions of the Hamiltonian.
This is a phase randomness assumption.  
The averaging process restores the 
translation invariance of the system on the average.  
The averaging method is clearly only appropriate when 
the number of impurities
is an extensive quantity.
  
It is convenient to rewrite the 
Hamiltonian in Eq.\ \ref{eq:himp} as
\begin{equation}
\hat{H} = \hat{H}_0 + \hat{V}_0 
+ \hat{V} - \hat{V}_0,
\end{equation}
where
$\hat{V}_0$ is the spatial average of $\hat{V}$.
We then define ${\cal G}_0(\vec{k},i \omega)$
as the unperturbed Green's function belonging to the 
Hamiltonian $\hat{H}_0 + \hat{V}_0$.
Both pieces of this Hamiltonian are diagonal in the momentum, and 
the second part gives only a rigid shift of the spectrum.
The perturbation is then $ \hat{V} - \hat{V}_0$,
which scatters electrons from a 
state $\vec{k}$ to a state $\vec{k}'$.  The scattering
amplitude is zero if
$\vec{k} = \vec{k}'$ because of the subtraction procedure.
It is important to subtract the average potential explicitly,
because the real part of the self-energy cannot be ignored
in this problem, as it often can be in other contexts.

The equation for the Green's function before averaging is
\begin{eqnarray}
{\cal G}(\vec{k},\vec{k}',i \omega)
& = & {\cal G}_0(\vec{k},i \omega) \delta_{\vec{k},\vec{k}'}
+ {\cal G}_0(\vec{k},i \omega) V_{\vec{k},\vec{k}'}
{\cal G}_0(\vec{k}',i \omega) \nonumber \\
& + & {\cal G}_0(\vec{k},i \omega) 
\sum_{\vec{k}''}V_{\vec{k},\vec{k}''}
{\cal G}_0(\vec{k}'',i \omega) V_{\vec{k}'',\vec{k}'}
{\cal G}_0(\vec{k}',i \omega) + \ldots
\nonumber \\
\label{eq:avgf}
\end{eqnarray}
Here 
\begin{equation}
V_{\vec{k},\vec{k}'} \equiv V 
\sum_i e^{i (\vec{k}-\vec{k}') \cdot \vec{R}_i}
- N_{imp} V \delta_{\vec{k},\vec{k}'}.
\end{equation}
It is evident that in the extreme low density limit
where the impurities have no influence on each other,
this equation will reduce to the single impurity case.
Quantities in Eq.\ \ref{eq:avgf}
are averaged using the prescription:
\begin{equation}
\overline{A} = \frac{1}{L^{3N_{imp}}} \prod_i \int d^3 R_i A,
\end{equation}
where $L$ is the linear dimension of the system (henceforth taken 
to be unity) and
$N_{imp}$ is the number of impurities.  This leads to
\begin{equation}
\overline{{\cal G}(\vec{k},\vec{k}',i \omega)}
={\cal G}_{imp}(\vec{k},i \omega) \delta_{\vec{k},\vec{k}'}.
\end{equation}
A Green's function diagonal in the momentum
${\cal G}_{imp}(\vec{k},i \omega)$
describes a state with uniform density.
This shows that the averaging procedure washes
out the density fluctuations which the impurities
induce in the ground state (and other states) of the
system.  Bound states 
are an example of such density fluctuations.
{\it Thus, there is no possibility of bound states   
in this approximation.}

Carrying out the averaging 
and neglecting diagrams with 
crossed lines \cite{agd} (an approximation which will
be discussed below) leads to the 
equation
\begin{eqnarray}
{\cal G}_{imp}(\vec{k},i \omega)
& = & {\cal G}_0(\vec{k},i \omega) 
+ {\cal G}_0(\vec{k},i \omega) 
n_{imp} V^2 \sum_{\vec{k}''}
{\cal G}_{imp}(\vec{k}'',i \omega) 
{\cal G}_{imp}(\vec{k},i \omega) \nonumber \\
&  + & 
{\cal G}_0(\vec{k},i \omega) 
n_{imp} V^3 \sum_{\vec{k}'}
{\cal G}_{imp}(\vec{k}'',i \omega)
\sum_{\vec{k}''}
{\cal G}_{imp}(\vec{k}'',i \omega) 
{\cal G}_{imp}(\vec{k},i \omega) + \ldots 
\label{eq:mimp}
\end{eqnarray}
The series is most conveniently summed by defining the
the self-energy
\begin{equation}
\Sigma(\vec{k},i \omega) 
= {\cal G}_0^{-1}(\vec{k},i \omega) 
- {\cal G}_{imp}^{-1}(\vec{k},i \omega) ,
\end{equation}
which leads to the equation
\begin{equation}
\Sigma(\vec{k},i \omega) 
= \frac{n_{imp} V^2 \sum_{\vec{k}'}
[i \omega - \epsilon_{\vec{k}'} - \Sigma(\vec{k},i \omega)]^{-1}}
{1 - V \sum_{\vec{k}'}
[i \omega - \epsilon_{\vec{k}'} - \Sigma(\vec{k},i \omega)]^{-1}}.
\label{eq:sig}
\end{equation}
It is then seen that $\Sigma$ is a function of frequency
alone for the short-range scattering potential.
For that reason, the equation is algebraic, not integral.   
In the Born approximation, the denominator
in this expression would be absent:
\begin{equation}
\Sigma(\vec{k},i \omega) 
= n_{imp} V^2 \sum_{\vec{k}'}
[i \omega - \epsilon_{\vec{k}'} - \Sigma(\vec{k},i \omega)]^{-1}
\label{eq:sborn}
\end{equation}

Eq.\ \ref{eq:sig} contains two physical
effects.  The band edges are moved inwards because
of level repulsion coming from the impurity potential.
This directly affects the real part of the self-energy.
The states are broadened because of the disorder which
means that $\vec{k}$-states are no longer eigenstates.
This directly affects the imaginary part of the self-energy.
Both of these effects tend to close the gap.

To estimate the critical value of the potential strength
at which this closure occurs, we must first
specify the model a bit more precisely.
Let us take the symmetric model of Fig.\ \ref{fig:symm}.
Then we need only determine when,
at midgap ( $\omega = 0$), the density of states
first becomes finite. 
At real frequencies, let us separate the real
and imaginary parts of the self-energy:
\begin{equation}
\Sigma(\omega) = \Sigma'(\omega) 
+ i \Sigma''(\omega).
\end{equation}  
Since we have
\begin{equation}
N(\omega) = - Im 
\sum_{\vec{k}} \frac{\Sigma''}
{(\omega-\epsilon_{\vec{k}}-\Sigma')^2 + 
\Sigma''^2},
\label{eq:nomega}
\end{equation}
we may simply increase the density $n_{imp}$ and scattering strength $V$
of the impurities,
two quantities which occur only in the combination
$n_{imp} V^2$, 
and determine when $ \Sigma''(0) \neq 0$.
First consider the Born approximation.
Symmetry dictates that 
$\Sigma'(0) = 0$.  The imaginary part of Eq.\ \ref{eq:sborn} 
at zero frequency is then
\begin{equation}
\Sigma''(0) = n_{imp} V^2 \Sigma''(0) f(0, \Sigma''(0)),
\label{eq:gapsg}
\end{equation}
where the function $f$ is defined as
\begin{equation}
f(\omega, \Sigma''(0)) \equiv N_0 \int 
\frac{d \epsilon}{(\omega-\epsilon)^2+(\Sigma''(0))^2},
\end{equation}
and the integral runs only over the energies for which the 
unperturbed density of states is
nonzero.
Eq.\ \ref{eq:gapsg} always has the solution $\Sigma''(0) = 0$.
It develops a second solution when
\begin{equation}
\frac{1}{n_{imp} V^2} =  f(0,0).
\end{equation}
$f(0, \Sigma''(0))$ is a positive,
monotonically decreasing function of
the nonnegative variable $\Sigma''(0)$.
This implies that the critical value of the disorder is
\begin{equation}
(n_{imp} V^2)_c = \frac{\Delta}{2 N_0}.
\end{equation}
This is when the inverse
relaxation time corresponds to the 
gap energy, as might be expected on physical grounds.

The consequences of band asymmetry are important for the
density of states.  Consideration of Eq.\ \ref{eq:sborn}
shows that the real part of the self-energy
is non-zero:
\begin{equation}
\Sigma(\omega = 0) \approx \frac{1}{\pi \tau} 
\log(\frac{\epsilon_u}{\epsilon_{\ell}}),
\end{equation}
and that the derivative of the derivative of the imaginary 
part also does not vanish at midgap.  Eq.\ \ref{eq:nomega}
then implies that the density of states
also has a nonzero slope at midgap.
I have computed numerically the solution
of Eqs.\ \ref{eq:sborn} and \ref{eq:nomega}
for the symmetric and asymmetric models.
The results are shown in Fig.\ \ref{fig:num}.
It is seen that the minimum in the density of 
states shifts away from the middle of the gap.
The chemical potential is given by a quite different
equation than that for the gap minimum.  It does
not coincide with the minimum.
 
The calculation for the full T-matrix equation,
Eq.\ \ref{eq:sig}, is only a little more complicated and will be omitted
here.  The result is that the threshold for 
complete closure of the gap is
unchanged, but the spectral weight in the gap is larger for the
same amount of disorder.  The conclusions about the symmetry of the
density of states in the gap are also
unchanged.

\subsection{Physical Picture}

It should be evident that there are profound
differences between the bound state calculations
and the impurity-averaging calculations.
The latter take into account only the potential fluctuations
and even the sign of $V$ is not very important.  In
the Born approximation, only $V^2$ enters the theory;
even when the T-matrix is used, the main effect is to 
alter the extent of the phase randomization, not to
create bound or antibound states.  The scattering perturbs 
and broadens the extended states.  The gap fills, but
the states which are in the
gap are extended states.  There is no question of impurity
band formation.  

In the bound state calculation,
the effect of the impurity potential on the extended
states is to give them a phase shift.  This does not
move the band edges.  The bound states are split off
from the bands.  The gap fills with localized states.

Very strong potential scattering in systems with a band
gap does lead to deep impurity levels.  There is no 
reason for the energy of these levels to be at
midgap.  This could only occur accidentally.
The decay of the impurity wavefunctions is 
very fast, as they are well split from the 
band states.  This prevents the formation
of impurity bands in all except pathological cases.

How may we combine the
results of both kinds of calculations,
the bound state and the impurity averaging ?
Let us first consider the set of all diagrams for
the single-particle Green's function 
${\cal G}(\vec{k},\vec{k}', \omega)$ associated with the
perturbation Hamiltonian $\sum_i V(\vec{r} - \vec{R}_i)$.
If the impurities are numbered from 1 up to N$_{imp}$,
then a diagram of n-th order perturbation 
corresponds to an ordered sequence of these integers
with n members.  A given integer may appear  
more than once (repeated scattering from a single impurity).
Intermediate momenta must be integrated over.  Each scattering
contributes a phase factor associated with the 
position of the impurity.  The T-matrix calculations correspond to
keeping sequences in which the same integer is repeated 
in succession many times.  If the impurities do not
interact, then only sequences containing a single integer
are considered.  This allows the bound state pole
to form.  The impurity averaging calculations omit the
phase factor and omit the 
momentum integration after the 
final appearance of any integer.  This is 
impurity averaging.  The neglect of
phase information prevents the build-up of
the bound-state pole.  A further approximation
is to discard sequences in which 
numbers are interleaved,
(an example is 1,2,1,2).  This is the noncrossing approximation.
It will be defined and discussed more carefully in connection with
calculations in the superconductiong state.

In the dilute limit,
the T-matrix approach is appropriate if $\omega$ lies in the
gap (and the unperturbed Green's function
decays exponentially in real space), 
and impurity averaging 
is appropriate if $\omega$ lies in one
of the bands.  It is therefore reasonable to
conclude that the DOS is properly given by
superposing the DOS from the two types of
calculations, with the proviso that there are $N - N_{imp}$
broadened band states and N$_{imp}$ gap states.  Here $N$ is the
total number of orbitals, proportional to the volume of the system.
As the impurity density increases and the
gap closes up, this energy separation argument
no longer works.  The continuum states overlap
in energy with the bound state.  The phase information
eventually becomes less important as the impurity
states mix with the continuum states.
The most likely scenario
as the density is increased seems to be as follows.    

For low concentrations, the bound states will
form a very narrow band in the gap.  The integrated
spectral weight in this band proportional to $n_{imp}$.
For subcritical disorder, there is 
no conduction associated with this band.  There is 
some spectral weight in the 
gap because of the broadening of the band states.
This weight is proportional to the total number of electrons,
not the number of impurities.  
However, it does not overlap in energy
with the bound state energy.
If the disorder is above critical, then
there is spectral weight everywhere in the
gap, because of broadening and shifting of the 
band states.  They overlap in energy with the 
bound state energy.  This broadens the bound states, 
turning them into resonances.  
The qualitative behavior for 
$n_{imp} V^2 < (n_{imp} V^2)_c$ and
$n_{imp} V^2 > (n_{imp} V^2)_c$
is shown in Figs.\ \ref{fig:sem1} and \ref{fig:sem2}, for the 
asymmetric case.

The resonances can, in principle, play a role
in conduction, since they are not necessarily localized.
However, their density will normally
be low compared to the band states, whose number is
proportional to the total number of sites, not the 
number of impurities.  This also means that if they do 
conduct, it is by hopping first into the continuum
and perhaps later onto another impurity, not directly
by impurity-impurity hopping as in an impurity band.

The question of localization of all these states
is subtle.  In the impurity-averaging method,
the single-particle Green's function is independent of 
position, seemingly indicating extended states.
However, when transport properties are calculated
using this method,  localization may appear in spite
of this.  Thus the states in the gap whose number  
of states is proportional to the total volume may or may
not be Anderson-localized by the disorder - the 
impurity-averaging
calculation of the single-particle Green's function
as carried out here gives no useful
information about this.  Conventional wisdom tells us that,
in the low impurity concentration regime considered here,
states near the chemical potential should be extended in three
dimensions and localized, but with extremely long localization lengths,
in two dimensions.

\section{s-wave superconductors}
\label{sec:swave}

The calculations of the effects of impurities in superconductors
are very analogous to the calculations in semiconductors.
This analogy is most easily exploited if we introduce the Nambu operators.
The defining equation is
\begin{equation}
\Psi_{\vec{k}}^{\dagger} = 
(c^{\dagger}_{\vec{k} \uparrow}, c_{-\vec{k} \downarrow}).
\end{equation}
The Pauli matrices 
$\tau_1,\tau_2,\tau_3$ and the identity matrix $\tau_0$ (omitted when clarity
requires), act in this two-dimensional space.  
The unperturbed mean-field Hamiltonian for
a superconductor with a constant gap $\Delta$ is:
\begin{equation}
\hat{H}_0 = \sum_{\vec{k}} \Psi^{\dagger}_{\vec{k}} \epsilon_{\vec{k}}
\tau_3 \Psi_{\vec{k}} +  
\sum_{\vec{k}} \Psi^{\dagger}_{\vec{k}} \Delta
\tau_1 \Psi_{\vec{k}},
\end{equation}
and the impurity potential is:
\begin{equation}
\hat{V} = V \sum_{\vec{k},\vec{k}'} \Psi^{\dagger}_{\vec{k}'}
\tau_3 \Psi_{\vec{k}}.
\end{equation}
The matrix Green's function is defined as
\begin{equation}
{\cal G}(\vec{k}, i \omega) =
- < T_{\tau} \Psi_{\vec{k}}(\tau) \Psi^{\dagger}_{\vec{k}}(0)>.
\end{equation}
In comparison to ordinary notation we find
\begin{eqnarray}
{\cal G}_{11}(\vec{k}, i \omega)  & = &  {\cal G}_{\uparrow \uparrow} (\vec{k}, i \omega)\\
{\cal G}_{22}(\vec{k}, i \omega)  & = & - {\cal G}_{\downarrow \downarrow}(\vec{k}, - i \omega)\\
{\cal G}_{12}(\vec{k}, i \omega)  & = & {\cal F}(\vec{k}, i \omega).
\end{eqnarray}
Since the density of states is independent of the spin direction
in singlet superconductors, we have that
\begin{equation}
N(\omega) = - \frac{1}{\pi} Im~Tr~{\cal G}_{11}(\vec{k}, \omega + i \delta).
\end{equation}
 
We may now easily calculate the unperturbed Green's function:
\begin{equation}
{\cal G}_0(\vec{k}, i \omega) = \frac{1}
{i \omega - \epsilon_{\vec{k}} \tau_3 - 
\Delta \tau_1} = 
- \frac{i \omega +  \epsilon_{\vec{k}} \tau_3 +
\Delta \tau_1}{\omega^2 +  \epsilon_{\vec{k}}^2 +
\Delta^2 }.
\end{equation}
The T-matrix is again defined by:
\begin{equation}
{\cal G}(\vec{k},\vec{k}', i \omega)  = {\cal G}_0(\vec{k}, i \omega) 
                                \delta_{\vec{k},\vec{k}'}
                                + {\cal G}_0(\vec{k}, i \omega) {\cal T}(i \omega) 
                                   {\cal G}_0(\vec{k}', i \omega),                                           
\end{equation}
which leads to:
\begin{equation}
{\cal T}(i \omega) = V \tau_3 + V^2 \tau_3 g_0(i \omega) \tau_3 +
                V^3 \tau_3 g_0(i \omega) \tau_3 g_0(i \omega) \tau_3 + \ldots,
\end{equation}
where
\begin{equation}
g_0(i \omega) = \sum_{\vec{k}}{\cal G}_0(\vec{k}, i \omega) = 
- \sum_{\vec{k}}\frac{i \omega \tau_0 +  \epsilon_{\vec{k}} \tau_3 +
\Delta \tau_1}{\omega^2 +  \epsilon_{\vec{k}}^2 +
\Delta^2 }.
\end{equation}
We shall assume particle-hole symmetry so that
\begin{equation}
\int^{\omega_c}_{-\omega_c} \epsilon_{\vec{k}} = 0,
\end{equation}
where $\omega_c$ is the cutoff frequency of the interaction,
which is much smaller than the bandwidth.  We also recall 
that in weak coupling
$\Delta = 0$ for $|\epsilon_{\vec{k}}| > \omega_c$.
Then:
\begin{eqnarray}
g_0(i \omega) & = & - (i \omega \tau_0 + \Delta \tau_1) N_0(\epsilon_F)
\int^{\omega_c}_{-\omega_c}\frac{d \epsilon }{\omega^2 +  \epsilon^2 +
\Delta^2 } \nonumber \\
& & - 
\int_{|\epsilon| > \omega_c}\frac{(i \omega \tau_0 + \epsilon \tau_3) 
N_(\epsilon) d \epsilon }{\omega^2 +  \epsilon^2 } \\ 
& = & - (i \omega \tau_0 + \Delta \tau_1) N_0(\epsilon_F)
\int^{\omega_c}_{-\omega_c}\frac{d \epsilon }{\omega^2 +  \epsilon^2 +
\Delta^2 } +  A_s \\
& = & - (i \omega \tau_0 + \Delta \tau_1)
\frac{\pi N_0(\epsilon_F)}{\sqrt{\omega^2 + \Delta^2}} + A_s.
\label{eq:asym}
\end{eqnarray}
The asymmetry factor is defined as
\begin{equation}
A_s \equiv \tau_3 N_A = 
\tau_3 \left[ \sum_{\vec{k}} \frac{\epsilon_{\vec{k}}}
        {\omega^2 + \epsilon_{\vec{k}}^2}\right], 
\end{equation}
where the sum runs over wavevectors whose energies are farther from the
Fermi surface than the cutoff frequency $\omega_c$.  We find
\begin{equation}
N_A = N_{u} \log(\frac{\epsilon_u}{\omega_c}) - 
N_{\ell} \log(\frac{\epsilon_{\ell}}{\omega_c}),       
\end{equation}
in the model of constant density of states beyond $\omega_c$.
$A_s$ also has an imaginary part, but it is
smaller by a factor of $\omega / \omega_c$.

We may now compute the T-matrix:
\begin{equation}
{\cal T}(i \omega) = V \tau_3 + V^2 \tau_3 ( a \tau_0 + b\tau_1 + N_A \tau_3) \tau_3
+ V^3 \tau_3 ( a \tau_0 + b\tau_1 + N_A \tau_3) \tau_3 
( a \tau_0 + b\tau_1 + N_A \tau_3) \tau_3 
+ \ldots ,
\end{equation}
where
the definitions $a(i \omega) =
- i  \pi N_0 (\epsilon_F) 
\omega / \sqrt{\omega^2 + \Delta^2}$, and
$ b(i \omega) = - \pi N_0 (\epsilon_F) 
\Delta / \sqrt{\omega^2 + \Delta^2}$ have
been made.
It remains to sum the geometric series and invert the 
resulting matrix:
\begin{eqnarray}
{\cal T}(i \omega) & = &  
V \tau_3 \sum_{n=0}^{\infty} 
[V ( a \tau_0 + b\tau_1 + N_A \tau_3) \tau_3]^n \nonumber \\
& = & V \tau_3 
[\tau_0 - V a \tau_3 - V b \tau_1 \tau_3 - V N_A \tau_0]^{-1} \nonumber \\
& = & \tilde{V} \tau_3 
[\tau_0 - \tilde{V} a \tau_3 - \tilde{V} b \tau_1 \tau_3 ]^{-1} \nonumber \\
& = & \frac{a  \tilde{V}^2 \tau_0 - b \tilde{V}^2 \tau_1 + \tau_3 \tilde{V}}
{1 - a^2 \tilde{V}^2 + b^2 V^2}.
\label{eq:tmat}
\end{eqnarray}
In these equations, we have made the definition, as before:
\begin{equation}
\tilde{V} = \frac{V}{1 + N_A V}.
\end{equation}
This shows that the binding potential is renormalized by 
the band asymmetry just as in the semiconductor case.
This limits its strength to
something on the order of the bandwidth.

Continuing to the real axis, we find
\begin{eqnarray}
T( \omega) & = & 
\frac{a(\omega)  \tilde{V}^2 \tau_0 - b(\omega) \tilde{V}^2 \tau_1 
+ \tau_3 \tilde{V}}
{1 - a(\omega)^2 \tilde{V}^2 + b(\omega)^2 V^2} \\
& = & \frac{a(\omega)  \tilde{V}^2 \tau_0 - b(\omega) 
\tilde{V}^2 \tau_1 + \tau_3 \tilde{V}}
{1 + (\pi N_0(\epsilon_F) \tilde{V})^2}.
\end{eqnarray}
This function is nonsingular, indicating that there 
are no bound states in the gap.
This arises from a cancellation of the frequency dependence
in the denominator, essentially that of
Anderson's theorem.

The asymptotic dependence of the Green's functions at 
large distances is of interest for later considerations.
The computational method is very similar for all cases,
so it is given in some detail here and abbreviated later.
In three dimensions, we have:
\begin{eqnarray}
{\cal G}(\vec{r}, i \omega) & = & \int e^{i \vec{k} \cdot \vec{r}}
{\cal G}(\vec{k}, i \omega) d^3 k \\ 
& = & - \frac{N_0}{4 \pi} \int d \epsilon \int_{-1}^1 dx \int_0^{2 \pi}
d \phi   \frac{i \omega \tau_0 +  \epsilon_{\vec{k}} \tau_3 +
\Delta_{\vec{k}} \tau_1}{\omega^2 +  \epsilon_{\vec{k}}^2 +
\Delta_{\vec{k}}^2 } e^{ikrx} .
\end{eqnarray}

The x-integral is performed by choosing a contour which
runs from $z = -1 + i \infty$ to $-1$, then from $-1$ to $+1$, then
from $+1$ to $1 + i \infty$.  Since we will use this contour several times,
it is shown in Fig.\ \ref{fig:int} for reference.
Let $ u \equiv (k_F + \epsilon / v_F) r$.
This variable must be treated carefully as the approximation of
linear dispersion is a very limited one.  We may extend the limits 
of integration over $u$ only if the integral is rapidly convergent.
This usually means that $u$ must be integrated last.
Bearing this in mind we have:
\begin{equation}
\oint d z e^{i u z} = \int^1_{-1}  dx  e^{i u x}
+ i \int_0^{\infty}  d y e^{i u(1 + i y)}
- i \int_0^{\infty}  d y e^{i u(-1 + i y)}, 
\end{equation}
as long as $u > 0$.  The contour integral vanishes because the function is 
analytic, so we find
\begin{eqnarray}
\int^1_{-1}  dx  e^{i u x} & = & -i e^{iu} \int_0^{\infty} e^{-uy} dy
+ i \int_0^{\infty} e^{-uy} dy \nonumber \\
  & = & - \frac{i}{u} (e^{iu} - e^{-iu}) \approx  
- \frac{i}{k_F r} (e^{iu} - e^{-iu}).
\end{eqnarray}
Substituting yields
\begin{eqnarray}
{\cal G}(\vec{r}, i \omega) & = & 
\frac{i N_0}{2 k_F r} \int d \epsilon 
[e^{i r (k_F + \epsilon/v_F)} - e^{-i r (k_F + \epsilon/v_F)}]
\frac{i \omega \tau_0 +  \epsilon \tau_3 +
\Delta \tau_1}{\omega^2 +  \epsilon^2 +
\Delta^2 } \nonumber \\
& = & \frac{i \pi N_0 e^{-dr/v_F}}{2 k_F r d}
[ (i \omega \tau_0 + i d \tau_3 + \Delta \tau_1) e^{i k_F r}
-  (i \omega \tau_0 - i d \tau_3 + \Delta \tau_1) e^{- i k_F r} ].
\end{eqnarray}
Here $d \equiv \sqrt{\omega^2 + \Delta^2}$.
As $\Delta \rightarrow 0$, this becomes
\begin{equation}
{\cal G}(\vec{r}, i \omega, \Delta = 0)
= - \frac{ \pi N_0 }{2 k_F r } e^{-|\omega| r /v_F}
[( sgn(\omega) \tau_0 +  \tau_3) e^{i k_F r} - 
( sgn (\omega) \tau_0 -  \tau_3) e^{-i k_F r}].
\end{equation}
This is the normal state Green's function.  The off-diagonal components are 
zero, and the diagonal one may be written as:
\begin{equation}
{\cal G}_{11}(\vec{r}, i \omega, \Delta = 0)
= - \frac{ \pi N_0 }{k_F r } e^{-|\omega| r /v_F}
e^{i k_F r~sgn(\omega)}.
\end{equation}
For future reference, the corresponding result in two dimensions is
\begin{equation}
{\cal G}_{11}(\vec{r}, i \omega, \Delta = 0)
= - \frac{N_0 }{\sqrt{2 \pi k_F r} } e^{-|\omega| r /v_F}
e^{i k_F r~sgn(\omega)}.
\end{equation}

At low frequencies, $\omega << \Delta$, we find
\begin{equation}
{\cal G}(\vec{r}, i \omega, \omega << \Delta)
= - \frac{ \pi N_0 }{2 k_F r } e^{- \Delta r /v_F}
[(sgn~(\omega) \tau_1 +  \tau_3 ) e^{i k_F r} 
- ( sgn~(\omega) \tau_1 - \tau_3 ) e^{-i k_F r}].
\end{equation}
The particle-hole part in the low frequency limit is
\begin{equation}
{\cal G}_{11}(\vec{r},  \omega, \omega << \Delta)
= - \pi N_0 e^{- \Delta r /v_F} \frac{\cos(k_F r)}{ k_F r } .
\end{equation}

The density of states is zero in the gap,
so the Greeen's function is purely
real.  The exponential damping in real space
is also due to this fact, the 
pole of the function
\begin{equation}
\frac{1}{\epsilon^2 + \omega^2 + \Delta^2}
\end{equation}
being off the real axis.
The decay length is the coherence length $v_F/\Delta$.

\section{Unconventional Superconductors}
\label{sec:dwave}

\subsection{Bound state energies}

I will restrict the discussion to the singlet case.
The new feature in the equations is that the
gap function is now $\vec{k}$-dependent,
and satisfies:
\begin{equation}
\sum_{\vec{k}} \Delta_{\vec{k}} = 0.
\end{equation}
The Green's function is the same as above,
and the T-matrix is still given by
\begin{equation}
{\cal T}(i \omega) = \tilde{V} \tau_3 + \tilde{V}^2 \tau_3 g_0(i \omega) \tau_3 +
                \tilde{V}^3 \tau_3 
                g_0(i \omega) \tau_3 g_0(i \omega) \tau_3 + \ldots,
\end{equation}
where 
\begin{eqnarray}
g_0(i \omega) & = & \sum_{\vec{k}} 
{\cal G}_0(\vec{k}, i \omega) \nonumber \\
& = & - i \omega N_0 \tau_0 
\frac{1}{4 \pi}\int d \Omega_{\vec{k}}
\left[\int^{\omega_c}_{-\omega_c} \frac{d \epsilon }
{\omega^2 +  \epsilon^2 + \Delta_{\vec{k}}^2 }+ 
\int_{|\epsilon| > \omega_c}  \frac{d \epsilon }
{\omega^2 +  \epsilon^2 } \right]
\label{eq:gn} \\
& = & - i \omega \tau_0 
\pi N_0 \frac{1}{4 \pi} \left[ \int \frac{d \Omega_{\vec{k}}}
{\sqrt{\omega^2 + \Delta_{\vec{k}}^2}} + A_s \right].
\end{eqnarray}
Here $\Omega_{\vec{k}}$ is the solid angle on the sphere.
The band asymmetry is represented by the second term in 
the equation, and comes from the integral over
energies far from the Fermi surface.  
Its effect is the same as above, i.\ e.\ , 
it results in the replacement of 
the bare potential $V$ by $\tilde{V} = V/(1+N_AV)$.

The equations for the unconventional case are, to this point,
actually somewhat simpler than the s-wave case. 
The T-matrix is obtained from Eq.\ \ref{eq:tmat}
by setting
$b = 0$ and $a = g_0(i \omega):$
\begin{equation}
T(i \omega)  =  
\frac{g_0(i \omega)  \tilde{V}^2 \tau_0 + \tau_3 \tilde{V}}
{1 - g_0(i \omega)^2 \tilde{V}^2}.                  
\end{equation}
Component by component, we find
\begin{equation}
T_{11}(\omega)  =  
\frac{g_0(\omega)  \tilde{V}^2  + \tilde{V}}
{1 - g_0( \omega)^2 \tilde{V}^2} = \tilde{V} 
\frac{1 + \tilde{V} g_0(\omega)}{1 - \tilde{V}^2 g_0^2(\omega)} = \tilde{V} 
\frac{1}{1 - \tilde{V} g_0(\omega)},
\end{equation}
for the up-spin electrons, and
\begin{equation}
T_{22}( - \omega)  =  
\frac{ - \tilde{V}}{1 + \tilde{V} g_0(- \omega)},
\end{equation}
for the down spins.
The corresponding equations for the binding energies are:
\begin{equation}
Re~g_0(\omega_b) = \frac{1}{\tilde{V}},
\end{equation}
and 
\begin{equation}
Re~g_0(- \omega_b) = - \frac{1}{\tilde{V}}.
\end{equation}
Since, from Eq.\ \ref{eq:gn},
$Re~g_0(\omega)$ is odd (if we exclude the asymmetry term), 
these equations heave the 
same solution for the bound state
energy $\omega_b$, as indeed they must.

To get some insight into the equations, we compute the
T-matrix for the 'polar' case, which has a line of nodes
on the equatorial plane of the Fermi surface: 
$\Delta_{\vec{k}} = \Delta_0 \cos(\theta_{\vec{k}}).$
We have
\begin{eqnarray}
Im~g_0(\omega) & = &  - \pi \frac{N_0}{4 \pi} \int_{-\infty}^{\infty} 
d \epsilon \int d \Omega_{\vec{k}}
[ \delta(\omega - E_{\vec{k}}) + \delta(\omega + E_{\vec{k}})] \\
& = & - \pi N_0 \int_{-\infty}^{\infty} d \epsilon \int_0^1 dx
\delta(|\omega| - \sqrt{\epsilon^2 + \Delta^2_0 x^2}),
\end{eqnarray}
which gives
\begin{eqnarray}
Im~g_0(\omega) & = &  \frac{- |\omega| N_0 \pi^2}{2 \Delta_0},~~|\omega| 
< \Delta_0 \\
& = & \frac{- |\omega| N_0 \pi}{\Delta_0} \sin^{-1}(\frac{\Delta_0}{\omega}),
~~ |\omega| > 
\Delta_0.
\end{eqnarray}
Computing the real part:
\begin{eqnarray}
Re~g_0(\omega)  & = & - \frac{1}{\pi} \int \frac{Im~g_0(\omega') d \omega'}
{\omega - \omega'}\\
  & \approx &  \frac{N_0 \pi}{2 \Delta_0} \int^{\Delta_0}_{- \Delta_0}
\frac{|\omega'| d \omega'}{\omega - \omega'}
+ \frac{N_0 \pi}{2} [\int_{\Delta_0}^{\omega_c}
\frac{d \omega'}{\omega - \omega'} + \int^{-\Delta_0}_{-\omega_c}
\frac{d \omega'}{\omega - \omega'}] \\
  & = &  \frac{N_0 \pi \omega }{2 \Delta_0} 
       \log|\frac{\omega^2}{\Delta_0^2 - \omega^2}|
        - \frac{N_0 \pi}{2} \log|\frac{(\omega_c - \omega)(\Delta_0 + \omega)}
        {(\omega_c + \omega)(\Delta_0 - \omega)}| \\
  & \approx & - \frac{N_0 \pi \omega}{ \Delta_0} 
\left(1 - \frac{1}{2} 
\log\left|\frac{\omega^2}{\Delta_0^2 - \omega^2} \right| \right),
\end{eqnarray}
at small $\omega$.  The asymmetry term $A_s$ has not been 
explicitly included in this expression, since we may more conveniently
include it in the potential strength.  
[$A_s = - N_0 \log(\epsilon_u/\epsilon_{\ell})$
in the simplest model density of states.]
For 
other gap functions which give an even smaller
density of states at low energies, the slope of the 
real part is completely
dominated by the first term, and gives a generic result:
\begin{equation}
Re~g_0(\omega) \sim - \beta \frac{N_0  \omega}{ \Delta_0},
\label{eq:beta}
\end{equation}
where $\beta$ is a number of order one. 
The bound state energy is:
\begin{equation}
\omega_b = - \frac{\Delta_0}{\beta N_0 \tilde{V}},
\end{equation}
as long as $ N_0 \tilde{V} >> 1$.  This is the
more generic result from Eq.\ \ref{eq:beta}, which we will also
use below.
The logarithmic corrections
in the polar case are only important if, for accidental
reasons, $1 + N_A V$ is exponentially small.

Once again, in this case, 
the bound state energy does not go to zero
even in the unitarity limit.  This is very important
for the properties of the bound state 
wavefunctions, to which we turn next.

\subsection{Asymptotic Spatial Dependence}

At large distances from the impurity, the 
wavefunctions for bound states are
determined by
the asymptotics of the diagonal elements of the 
real-space Green's functions 
evauated at $\omega_b$ \ref{eq:psi}.
The off-diagonal elements are also of interest, as they 
represent the suppression of the gap in the 
neighborhood of the impurity.  Their asymptotic spatial dependence
is always the same as that of the diagonal elements, so I
do not give expressions for them separately.

In three dimensions I shall concentrate on the polar case.
The gap takes the form
$\Delta(\vec{k}) = \Delta_0 \cos \theta_{\vec{k}}$.
We shall take a spherical Fermi surface; for this 
situation the gap has a line of nodes around the equator.
As in the previous section, we assume that the 
pair wavefunction is a singlet in spin space.
This total pair wavefunction does not satisfy the Pauli
principle.  However, it is the simplest
example of an unconventional state 
with a line of nodes in three dimensions
and the calculations for it are already complicated.

Now take the direction $\hat{r}$ from the impurity to
be in the z-direction.  I will work out the Green's function
in the main text for this case.  For other cases, the 
intermediate steps are relegated to the appendix.

Then
\begin{equation}
{\cal G}(r \hat{z}, i \omega_b) =
- \frac{N_0}{2} \int d \epsilon \int_{-1}^1 e^{i u x}
\frac{i \omega_b + \epsilon \tau_3 + \Delta_0 x \tau_1}
{\omega_b^2 + \epsilon^2 + \Delta_0^2 x^2} dx.
\end{equation}
The important integral (because it expresses the
angular dependence) is the x integral.
\newpage
This is:
\begin{eqnarray}
I(\epsilon, i \omega_b) & = & \int_{-1}^1 e^{i u x}
\frac{i \omega_b + \epsilon \tau_3 + \Delta_0 x \tau_1}
{\omega_b^2 + \epsilon^2 + \Delta_0^2 x^2} dx \nonumber \\
  & =  & \frac{1}{\Delta_0} \int_{-1}^1 e^{i u x}
\frac{i \Omega + s \tau_3 + x \tau_1}
{\Omega^2 + s^2 +  x^2} dx\nonumber \\
  & = & \frac{1}{\Delta_0} \oint e^{i u z}
\frac{i \Omega + s \tau_3 + z \tau_1}
{\Omega^2 + s^2 +  z^2} dz \nonumber \\
  &  & - i e^{ i u } \frac{1}{\Delta_0} \int_0^{\infty}
        e^{ - u y}
\frac{i \Omega + s \tau_3 + (1 + i y) \tau_1}
{\Omega^2 + s^2 +  (1 + iy)^2} dy \nonumber \\
  &  & + i e^{ - i u } \frac{1}{\Delta_0} \int_0^{\infty}
        e^{ - u y}
\frac{i \Omega + s \tau_3 + (-1 + iy) \tau_1}
{\Omega^2 + s^2 +  (-1 + iy)^2} dy.
\end{eqnarray}
The method is the same as in the s-wave case, and 
we use the same contour, as shown in Fig.\ \ref{fig:int}.
The scaled variables $s = \epsilon/\Delta_0$,
$\Omega = \omega_b/\Delta_0$, have been used, and again
$u = (k_F + \epsilon/ v_F) r $.  We are interested in the limits
$\Omega << 1 $, and $ k_F r >> 1.$  The contour integral is easily
performed by the residue theorem, and the strongly peaked
function $e^{-u y }$ allows us to do the other two integrals.
Thus,
\begin{equation}
I(\epsilon, i \omega_b)  =  \frac{\pi }{\Delta_0}
e^{- u \sqrt{\Omega^2+s^2}}  
\frac{i \Omega + s \tau_3 + i \sqrt{\Omega^2 + s^2} \tau_1}
{\sqrt{\Omega^2 + s^2}} - i \frac{e^{iu}}{u \Delta_0} 
\frac{s \tau_3 + \tau_1}{1 + s^2 + \Omega^2}
+ i \frac{e^{-iu}}{u \Delta_0} 
\frac{s \tau_3 - \tau_1}{1 + s^2 + \Omega^2}.
\end{equation}

For $u >>1$, we have the useful integrals:
\begin{equation}
I_1 \equiv \int_{-\infty}^{\infty}
\frac{e^{-u(\Omega^2 + s^2)^{1/2}}ds}{(\Omega^2 + s^2)^{1/2}}
\approx (\frac{2 \pi}{u |\Omega|})^{1/2} e^{-u |\Omega|},
\end{equation}
\begin{equation}
I_2 \equiv \int_{-\infty}^{\infty}
e^{-u(\Omega^2 + s^2)^{1/2}} ds
\approx (\frac{2 \pi |\Omega|}{u})^{1/2} e^{-u |\Omega|}.
\end{equation}
This yields
\begin{eqnarray}
\int_{-\infty}^{\infty} ds I(\epsilon,\omega_b) & = & 
\frac{\pi}{\Delta_0}  e^{-u |\Omega|}
\left[ i \Omega (\frac{2 \pi}{u |\Omega|})^{1/2} 
+ i \tau_1 (\frac{2 \pi |\Omega|}{u})^{1/2} \right] \nonumber \\
& + & \frac{\pi(\Omega + \sqrt{1 + \Omega^2} \tau_3 -i \tau_1)}
{\Delta_0 k_F r \sqrt{1 + \Omega^2}} 
\exp[i k_F r - \sqrt{1 + \Omega^2} \Delta_0 r / v_F] \nonumber \\
& + & \frac{\pi(- \Omega + \sqrt{1 + \Omega^2} \tau_3 -i \tau_1)}
{\Delta_0 k_F r \sqrt{1 + \Omega^2}} 
\exp[- i k_F r - \sqrt{1 + \Omega^2} \Delta_0 r / v_F].
\end{eqnarray}
The Green's function at large distances is therefore
\begin{eqnarray}
{\cal G}(r \hat{z}, i \omega_b )& = & - \frac{N_0 \Delta_0}{2}
\int_{-\infty}^{\infty} ds I(\epsilon,\omega_b)  \nonumber \\
  & = & \frac{-i \pi N_0}{2}  (\frac{2 \pi |\omega_b|}
  {\Delta_0 k_F r})^{1/2}
  e^{-k_F r |\omega_b| / \Delta_0} (sgn~(\omega_b) + \tau_1)\nonumber \\
  & & -  \frac{\pi N_0 e^{-(\omega_b^2 + \Delta_0^2)^{1/2}r/v_F}}
  {2 (\omega_b^2 + \Delta_0^2)^{1/2}} \times \nonumber \\
  & & [ (\omega_b + (\omega_b^2 + \Delta_0^2)^{1/2} \tau_3 - i \tau_1) 
  e^{i k_F r} + ( - \omega_b + 
  (\omega_b^2 + \Delta_0^2)^{1/2} \tau_3 - i \tau_1) 
  e^{-i k_F r}].
\end{eqnarray}  

The component of most interest is the particle-hole part
at real frequencies:
\begin{eqnarray}
{\cal G}_{11}(r \hat{z}, \omega_b)
  & = &- \frac{\pi N_0}{2}  (\frac{2 \pi | \omega_b|}  
         {\Delta_0 k_F r})^{1/2}
  e^{-k_F r | \omega_b| / \Delta_0}~sgn~(\omega_b) \nonumber \\
  &+ & \frac{i \pi N_0}{2} 
     e^{-(\Delta_0^2 - \omega_b^2)^{1/2}r/v_F} \times \nonumber \\
  & & \left\{ \frac{\omega_b + (\Delta_0^2 - \omega_b^2 )^{1/2}}
     {(\Delta_0^2 - \omega_b^2 )^{1/2}}  
      e^{i k_F r} 
   - \frac{ \omega_b - (\Delta_0^2 - \omega_b^2 )^{1/2}}
     {(\Delta_0^2 - \omega_b^2 )^{1/2}}  
     e^{-i k_F r} \right\}.
\label{eq:ph1}
\end{eqnarray}  

The first term is the pole contribution, which arises from
interference of the low energy states near the line of 
nodes.  If the state is precisely at the chemical potential,
then these states interfere to give a power law decay.
There is no oscillatory component because the states
on the nodal line have $\vec{k} \cdot \vec{r} = 0$
for this direction of $\vec{r}$.  The length scale
of this decay is $\Delta_0/|\omega_b|k_F$,
which is the interatomic spacing unless the
binding energy happens to be very small.

The second term is the
stationary-phase contribution 
which is due to the states near the north and south
poles of the Fermi surface.  Since these are gapped,
we get an exponential falloff which reflects the
energy gap at these points.  There is an oscillatory
behavior because $\vec{k} \cdot \vec{r} = \pm k_F r$
at these points. 
The length scale of the decay is the 'local'
coherence length (coherence length at the pole)
$\Delta(k \hat{z})/v_F$ 
if the binding energy is 
small.
 
A second special case of interest is when the direction
of $\vec{r}$ is on the plane of nodes.  Then we have
\begin{equation}
{\cal G}(r \hat{x}, i \omega_b) =
- \frac{N_0}{4 \pi} \int d \phi 
\int d s \int_{-1}^1 e^{i u x}
\frac{i \omega_b +  s \tau_3 + \cos\phi(1 - x^2)^{1/2} \tau_1}
{\omega_b^2 + s^2 + \cos^2\phi (1 - x^2)} dx.
\label{eq:app1}
\end{equation}

The details of the integration are given in the appendix.
The result is:
\begin{equation}
{\cal G}(r \hat{x}, i \omega_b) = \frac{- \pi N_0}{2 k_F r} e^{-|\omega_b|r/v_F}
        \left[ e^{i k_F r} (sgn~(\omega_b) \tau_0 + \tau_3) 
            -  e^{- i k_F r} 
        (sgn~(\omega_b) \tau_0  - \tau_3) \right] .
\label{eq:app2}
\end{equation}
The particle-hole part is
\begin{equation}
{\cal G}_{11}
(r \hat{x}, \omega_b) = 
\frac{- \pi N_0}{2 k_F r} e^{-|\omega_b|r/v_F}
        e^{i k_F r~sgn~(\omega_b) }.
\label{eq:ph2}
\end{equation}
These are essentially the normal state results.
The point here is that the stationary phase points at
$\vec{k} = \pm k_F \hat{x}$ have no gap and the behavior
is therefore entirely normal.

The general case is rather complicated.  Let the direction
from the impurity be inclined at an angle $\psi$
to the z-axis.  Let us define 
\begin{equation}
{\cal G}(r, \psi, i \omega_b)
\equiv {\cal G}(r (\sin \psi \hat{x} + \cos \psi \hat{z}, i \omega_b))
\end{equation}
We have that
\begin{equation}
{\cal G}(r, \psi, i \omega_b)
= - \frac{N_0}{4 \pi}
\int_0^{2 \pi} d \phi \int d \epsilon \int_{-1}^1 dx
e^{iux} \frac{i \omega_b + \epsilon \tau_3 + \Delta_0 \tau_1
(ax - b \sqrt{1 - x^2})}{\omega_b^2 + \epsilon^2 + \Delta_0^2
(ax - b \sqrt{1 - x^2})^2},
\label{eq:app3}
\end{equation}
where $a \equiv \cos \psi, b = \cos \phi \sin \psi$,
and $u \equiv (k_F + \epsilon /v_F)r$.
The coordinates for the integration have been rotated so that
the polar axis defined by $x = \pm 1$ is along $\vec{r}$.
The particle-hole part of the Green's function 
is found in the appendix to be: 
\begin{eqnarray}
{\cal G}_{11}(r, \psi, \omega_b )
 & = &   \frac{\pi N_0 sgn~(\omega_b)}{k_F r} 
    \left( \frac{ 2 | \omega_b|}
    { \Delta_0 \cos^2 \psi \sin^2 \psi} \right)^{1/2}  \sin ( k_F r \sin \psi)  
  e^{- k_F r |\omega_b| |\cos \psi|/ \Delta_0} \nonumber \\
  &  &  - \frac{i N_0 \pi}{2 k_F r} 
        \left( e^{i k_F r}\frac{\omega_b + \sqrt{\Delta_0^2 \cos^2 \psi - \omega_b^2}}
                {\sqrt{\Delta_0^2 \cos^2 \psi - \omega_b^2}} 
         -  e^{-i k_F r}\frac{\omega_b - \sqrt{\Delta_0^2 \cos^2 \psi - \omega_b^2}}
                {\sqrt{\Delta_0^2 \cos^2 \psi - \omega_b^2}} \right)\nonumber \\
            \times e^{-\sqrt{\Delta_0^2 \cos^2 \psi - \omega_b^2}r/v_F}.
\label{eq:ph3}
\end{eqnarray}
This general form does not reduce to Eq.\ \ref{eq:ph2}
when $\psi = \pi/2$ or to 
Eq.\ \ref{eq:ph1} when $\psi = 0$.  It 
is only valid in the intermediate regime 
of angles.  
The result does, however, show that the two distinct decay behaviors
noted above are both present in the generic case.
  
\subsection{Two dimensions}                                                      
We now turn to the case most relevant to high-T$_c$, which is
two dimensions with a singlet d-wave gap, (which is consistent with the
Pauli principle).  
The Fermi surface is a circle and the
gap will be taken as
\begin{equation}
\Delta(\vec{k}) = \Delta_0 \cos 2 \phi_{\vec{k}}.
\end{equation}
This d-wave gap has nodes at the intersection of the 
lines $k_x = \pm k_y$ with the Fermi surface.
Let us first determine the Green's function
for the direction along one of the axes.
We have 
\begin{equation}
{\cal G}(r \hat{x}, i \omega_b ) = - \frac{2 N_0}{\pi}
\int d \epsilon \int_{-1}^1 \frac{dv}{\sqrt{1-v^2}}
\frac{i \omega_b + \epsilon \tau_3 + \Delta_0 \tau_1 (2 v^2 -1)}
{\omega_b^2 + \epsilon^2 + \Delta_0^2 (2 v^2 -1)^2} e^{iuv}.
\label{eq:2dgx1}
\end{equation}
Here we have defined $v = \cos \phi_{\vec{k}}$ 
so that $ \Delta = \Delta_0 (2v^2-1)$.
Also $u = (k_F + \epsilon/v_F) r$.  
The integrations are performed in the appendix.  The 
particle-hole part is
\begin{eqnarray}
{\cal G}_{11}(r \hat{x}, \omega_b ) 
& = & - \frac{2 i N_0}{\pi} \times \nonumber \\
  &  & \left\{ \pi  
  (\frac{(\sqrt{\Delta_0^2 - \omega_b^2} - \Delta_0 )^{1/4} \Delta_0^{1/4}}
{(\Delta_0^2 - \omega_b^2)^{1/4}}) (\frac{8 \pi }{k_F r})^{1/2} 
\exp\left(\frac{-k_F r}{2 \sqrt{\Delta_0}}
\sqrt{\sqrt{\Delta_0^2 - \omega_b^2}-\Delta_0}\right) \right.
 \times  \nonumber \\
& & i ~sgn~(\omega_b) \cos\left(\frac{k_F r}{2 \sqrt{\Delta_0}}
\sqrt{\sqrt{\Delta_0^2 - \omega_b^2}+\Delta_0}\right) \nonumber \\
  &  & + \frac{(1-i)\pi}{2} \sqrt{\frac{\pi}{k_Fr}} e^{ik_Fr} 
e^{-\sqrt{\Delta_0^2 - \omega_b^2} r/v_F}
\frac{i \Omega + i \sqrt{\Delta_0^2 - \omega_b^2}}
{\sqrt{\Delta_0^2 - \omega_b^2}},\nonumber \\
& &  + \frac{(1-i)\pi}{2} \sqrt{\frac{\pi}{k_Fr}} e^{-ik_Fr} 
\left. e^{-\sqrt{\Delta_0^2 - \omega_b^2} r/v_F}
\frac{i \Omega - i \sqrt{\Delta_0^2 - \omega_b^2}}
{\sqrt{\Delta_0^2 - \omega_b^2}} \right\}.   
\label{eq:giwrx}
\end{eqnarray}
The peculiar phase factors reflect the orientation of the 
nodal directions relative to the crystal axes.
The overall behavior is very similar to the three-dimensional case,
Eq.\ \ref{eq:ph1}

The direction along the node is more easily computed.
\begin{equation}
{\cal G}(r \frac{\hat{x} + \hat{y}}{\sqrt{2}}, i \omega_b) = - \frac{2 N_0}{\pi}
\int d \epsilon \int_{-1}^1 \frac{dv}{\sqrt{1-v^2}}
\frac{i \omega_b + \epsilon \tau_3 + \Delta_0 \tau_1 v \sqrt{1-v^2}}
{\omega_b^2 + \epsilon^2 + 2 \Delta_0^2 v^2(1- v^2)} e^{iuv},
\label{eq:2dgx2}
\end{equation}
where the part proportional to $\tau_1$
vanishes by symmetry.  
The integrations are performed in the appendix, with the result that:
\begin{eqnarray}
{\cal G}_{11}(r \frac{\hat{x} + \hat{y}}{\sqrt{2}}, \omega_b ) &  
= & - \frac{2 i N_0}{\pi} \times \nonumber \\
  &  & \left[ (\frac{4 \pi^3 \sqrt{2}}{k_F r})^{1/2}
\frac{~sgn~(\omega_b)}{(\Delta_0^2 - \omega_b^2)^{1/4}}
(\sqrt{\Delta_0^2 - \omega_b^2}- \Delta_0)^{1/4} \Delta_0^{1/4}
e^{-k_F r  (\sqrt{\Delta_0^2 - \omega_b^2}- 
\Delta_0)^{1/2}/\sqrt{2 \Delta_0}} \right. \nonumber \\
  &  & + \frac{(1-i) \pi}{2}~\frac{e^{ik_Fr}}{\sqrt{k_Fr}}~
  (~sgn~(\omega_b) + 1 )
e^{- |\omega_b| r/v_F}\nonumber \\                                           
  &  &  \left. + \frac{(1+i) \pi}{2}~\frac{e^{-ik_Fr}}
         {\sqrt{k_Fr}}~( ~sgn~(\omega_b) - 1 )
e^{- |\omega_b| r/v_F} \right] .
\label{eq:grxy}
\end{eqnarray}
This is essentially the two-dimensional normal state 
result.

Now consider the general case, where the direction is
inclined at an angle $\psi$ from the diagonal
and define:
\begin{equation}
{\cal G}(r, \psi, i \omega_b )
\equiv  {\cal G}(i \omega_b, r \cos(\frac{\pi}{4} - \psi)\hat{x} + 
\sin(\frac{\pi}{4} - \psi) \hat{y}). 
\end{equation}
The result, derived in the appendix, is:
\begin{eqnarray}
{\cal G}_{11}(r, \psi, \omega_b )
&  = & - \frac{2 i N_0}{\pi} \times \nonumber \\
& & \left\{
 (\frac{8 \pi^3 \sqrt{2}}{k_F r |\cos \psi|})^{1/2} \right.
\frac{sgn~\omega_b }{(\Delta_0^2 -\omega_b^2)^{1/4}}  
(\sqrt{\Delta_0^2 -\omega_b^2}-\Delta_0)^{1/4} \Delta_0^{1/4}
\times \nonumber \\
& & e^{-k_F r  |\cos \psi| 
(\sqrt{\Delta_0^2 -\omega_b^2}-\Delta_0)^{1/2}/\sqrt{2 \Delta_0}}
\cos \left[i k_F r \sin \psi 
(\sqrt{\Delta_0^2 - \omega_b^2}-\Delta_0)^{1/2}/
\sqrt{2 \Delta_0}\right] \nonumber \\
& + & 
\frac{(1-i) \pi}{2}~\frac{e^{ik_Fr \cos \psi}}
{\sqrt{k_Fr|\cos \psi|}}~(sgn~\omega_b +1)
e^{- |\omega_b \cos \psi| r/v_F} \nonumber \\
& + &
\left. \frac{(1+i) \pi}{2}~\frac{e^{-ik_Fr \cos \psi}}
{\sqrt{k_Fr |\cos \psi|}}~(sgn~\omega_b - 1 )
e^{- |\omega_b \cos \psi| r/v_F}] \right\}.
\label{eq:grpsi}
\end{eqnarray}

\subsection{Discussion}

As compared with the semiconductor, there are several
interesting differences in the bound state
wavefunctions of the unconventional superconductor.
The most important is the fact that the decay length
may become longer, not shorter, as the bound state
energy approaches midgap.  In both cases, however, there
is nothing special about the middle of the gap
in a real system, so exponential decay is still the norm.

In the superconductor, the wavefunctions have two components
when viewed in real space.  The two components 
correspond to two different exponential decay lengths.
One of these lengths is determined by the gap along the direction
of propagation.  This length is $v_F/\sqrt{\Delta(\hat{k})^2-\omega_b^2}$. 
This is similar to the s-wave case except for the anisotropy.
Indeed, this contribution
comes from the fully gapped region of the Fermi surface.  
The second length is $v_F/|\omega_b \cos \psi|$, where
$\psi$ is the angle away from the nodal plane.
This contribution is peculiar to the unconventional
case, arising from the gap nodes.  
These lengths are anisotropic, so the wavefunctions
are also anisotropic, with 'arms' in the directions of the 
gap nodes.  Some very nice pictures of
these wavefunctions may be found in Ref.\ \cite{salkola}.

The decay is 
always exponential in all directions unless, for accidental reasons, the 
bound state energy is zero.

\section{Many impurities in superconductors}
\label{sec:many}

\subsection{Introduction}

The discussion of many impurities in the metal and 
the superconductor are usually considered to be parallel,
and the same equations, with only the generalization to the
Nambu formalism, are used for both \cite{agd}.  
In articular, the method of impurity averaging is 
not modified.  Hence, in this section we shall not repeat the 
calculations of Sec.\ \ref{sec:mase}.

For unconventional superconductors,
without a hard gap, this is fundamentally reasonable.  
We have seen that impurity averaging becomes valid when the
continuum states overlap the bound state energy.  The
constructive buildup of phase required to make the bound state
is destroyed when there is overlap with the continuum states, themselves
possessing a random shift.  This is always the case, as the bound state
energy is not in midgap.  Accordingly, the pole found in the
T-matrix calculations should immediately broaden into a resonance.
This being said, one should address four basic issues which
arise in practical calculations.  These are: 
symmetry of the DOS around the chemical potential, 
the nature of the states at the
chemical potential, the validity of the noncrossing approximation,
and impurity band formation.  

The first two are relatively easily dealt with.
The third and fourth are treated in the next two sections.

The density of states is not symmetric about the chemical potential, even when 
particle-hole symmetry is valid.  This has already been pointed recently
by other authors in the magnetic impurity case \cite{salkola}, \cite{flatte}.
This arises from the same source as in the 
semiconductor.  Symmetry of the DOS requires that the bands themselves be 
symmetric over their whole energy range, not just over the neighborhood of the 
chemical potential.  It appears that this fact is often not taken into account 
in practical calculations.  The classic reference \cite{agd} advises us to 
neglect the real part of the energy shift.  This is indeed safe for the 
neighborhood of the Fermi energy in a metal, and for weak scattering in s-wave 
superconductors, the cases discussed in Ref.\ \cite{agd}.  It is not valid in 
unconventional superconductors with a soft gap.  

The states at the chemical potential are sometimes termed 'bound 
states', and their heritage as the descendants of the T-matrix poles is 
emphasized.  It should be clear from the discussion that this is not correct.  
These states are the broadened and shifted continuum states.  The number of 
such states in any range of energies is proportional to the total number of 
orbitals in the system.  The daughters of the bound 
states will generally live in a 
resonance away from the chemical potential.  Their number is proportional to 
the number of impurities.  

\subsection{Noncrossing approximation}

One approximation used in nearly 
all calculations of superconducting properties
is the noncrossing approximation.  
This has been questioned in recent work, \cite{neresyan},
and I reproduce and expand this criticism here.
This approximation is defined
diagramatically by representing each impurity
as a cross through which momentum flows and is conserved.
Let us restrict the discussion in this section
to the Born approximation for simplicity.
Then two typical diagrams for the normal state
are shown in Fig.\ \ref{fig:non1}.  We take a circular Fermi curve
in a two-dimensional system.  Diagram (a) in Fig.\ \ref{fig:non1}
has no crossed lines, whereas diagram (b) does.
The same processes are shown in 
momentum space in the diagrams in Fig.\ \ref{fig:non2}.
Diagram (a) describes a retraceable
path, while diagram (b) contains a circuit.  Therefore, the
second diagram must satisfy one 
additional momentum conservation condition.  It is therefore
smaller in magnitude than the first.  Explicit calculation 
shows that the small parameter involved is $(1/k_F \ell)$,
where $\ell$ is the mean free path.  

In a two-dimensional d-wave superconducting state at low temperatures, 
this argument must be reconsidered.
All diagrams must effectively satisfy additional
momentum constraints, as only scattering between
gap nodes, situated at $(\pm k_F/\sqrt{2}, \pm k_F/\sqrt{2})$
is important.  The two diagrams in Fig.\ \ref{fig:non3}
again correspond to the processes of Fig.\ \ref{fig:non1}.  The two
diagrams are of roughly equal weight, even though the second one is crossed.
There are no {\it additional} constraints which must be satisfied by the 
second diagram.  

The noncrossing approximation is therefore very questionable 
in unconventional superconductors with point nodes.
The authors of Ref.\ \cite{neresyan}
attempt to go beyond this approximation
in a not very realistic model.  
No calculations of 
transport properties have been carried out except using 
the noncrossing approximation.
It certainly is difficult to justify for the d-wave states considered
in the context of high-temperature superconductivity, if these materials are
taken to be two-dimensional.

\subsection{Impurity band formation}

Does there exist the possibility of 
the formation of an impurity band ? 
Does conduction in this band influence, or
even dominate, the 
transport properties in the limit of low temperatures ?

We begin, as in Sec.\ \ref{sec:mase}, by considering two impurities.
The overlap now depends, to some extent,
on the direction of the vector connecting the two impurities,
with the direction of minimum gap being the 
direction of maximum overlap. 
If we take the two-dimensional example summarized in Eq.\ \ref{eq:grpsi}
the overlap proceeds according to $\exp(-k_F r |\cos \psi|)$
or $\exp(-|\omega_b \cos \psi| r / v_F)$.

The introduction of many impurities 
always brings one new number to the problem:
the average distance between impurities,
which shall be denoted by $\ell_{imp} \approx n_{imp}^{-1/d}$,
where $n_{imp}$ is the number of impurities per unit volume.
If the impurities are not
identical, we have a disorder parameter $W$, 
defined as the width of the 
distribution of the potential strength,
previously the single number $V$, of the 
impurities.  The usual model of impurities
is that they are all identical: $W=0$,
but we will consider also $W \neq 0$.
If interactions on the impurities
are important, we may introduce a Hubbard-type parameter
to describe the interaction strength.
We shall not discuss this possibility, but only note that
this also introduces a breaking of the band symmetry
which can move the bound state away from midgap.
This situation is treated in Ref.\ \cite{salkola}.

The first question for band formation is the following:
given a wavepacket located at an impurity site,
is it more likely to hop to a neighboring impurity,
or to leak into a continuum state ?  If the latter, then 
the impurities merely form a system of resonances
and averaging procedures should be approximately valid.
In this case, we may make arguments similar to those
for the semiconductor to argue that the results
of the two types of calculations may be combined.

If $W=0$, then the bound state energy $\omega_b$ is fixed
at some position relative to the chemical potential $\mu$.
In the general case $\omega_b \neq \mu$,
{\it even in the unitary limit}, as we have seen above.
This means that there is a finite density of states
at $\omega_b$.  The lifetime of the wavepacket for 
decay into the continuum $\tau_c$ is given by
\begin{equation}
\frac{1}{\tau_c} = \pi N(\omega_b) |V|^2. 
\end{equation}
Near the unitary limit, we have 
$|V| > 1/N_0$.  Using Eq.\ \ref{eq:beta},
we then find
\begin{equation}
\frac{1}{\tau_c} > \frac{\omega_b}{\Delta N_0}.
\end{equation}
This may be anomalous only if $|\omega_b| << \Delta_0$.
The rate for interimpurity hopping $1/\tau_i$, is
of order
\begin{equation}
\frac{1}{\tau_i} \sim \frac{e^{-\ell_{imp}/\xi_d}}{N_0},
\end{equation}
where $\xi_d$ is the {\it minimum} 
decay length.
As shown above, we have that $\xi_d = 
v_F/|\omega_b|$.

The criterion
for band formation is then
\begin{equation}
e^{-\ell_{imp}/\xi_d} \geq 
\   \frac{\pi \beta \omega_b}{\Delta}.
\end{equation}
Unless $\omega_b$ is accidentally very small,
this means that we must have $\ell_{imp} \sim \xi_d$.
However, this is the dirty limit.  This limit
does not exist for unconventional 
superconductors because the critical temperature $T_c$
is a sensitive
function of impurity concentration and the 
situation $\ell_{imp} \sim \xi_d$ 
corresponds to $T_c \rightarrow 0$.
Hence band formation does not occur in the 
$W=0$ case.  
The only possible exception would be 
if, by some accident, $|\omega_b| << \Delta$. 
This is an intriguing possibility.  However,
it is not related to the unitary limit. 

Now consider the case of finite $W$.  There
will be a distribution of bound state 
energies.  If this distribution does not
include the chemical potential, then the previous conclusion that no 
impurity band forms
remains valid, as no quasi-bound states have low enough
energy to be anything but broad resonances.
The interesting case is when the distribution
is broad enough that some of the bound 
states have very small $|\omega_b|$,
($|\omega_b| << \Delta$).

We may build up the state by considering pairs of impurities.
Almost all such pairs which involve a low-energy
impurity state (energy $|\omega_{b1}|)$
will then involve as the second member a 
state for which $|\omega_{b2}| \sim \Delta$.  
These states will mix, with overlap matrix element
$M_{12}$.  Under the influence of the 
mixing, state 1 then has probability amplitude on site 2
of $ M_{12}^2/(\omega_{b1} - \omega_{b2})^2$,
when $|M_{12}| << |(\omega_{b1} - \omega_{b2})|$.
The transition rate to the continuum for this state is
then 
\begin{equation}
\frac{1}{\tau_c} \sim \frac{M_{12}^2 \pi \beta}
{|\omega_{b2}| \Delta N_0}.  
\end{equation}
If site 1 has many such neighbors, then the 
transition rate to the continuum is multiplied by the
number of neighbors.  What happens is that the 
stae leaks first through an impurity and then into the continuum.
Rare transitions from one low energy impurity
state to another will therefore not lead to the
formation of a well-defined band, and transport
occurs through the extended states.

The result is a collection of resonances
together with a set of broadened 
continuum levels.  As in the semiconductor case,
the total number of continuum states is equal to 
the number of atomic orbitals 
in the sample; the number of resonances is
equal to the number of impurities.  The resulting total density of
states is shown in Fig.\ \ref{fig:smany}.
We must conclude that the low temperature behavior
in this system is never dominated by interimpurity hopping.
As a result, calculations using impurity averaging methods
should lead to correct results.  However, the noncrossing 
approximation may be dangerous in two dimensions, and
the use of a symmetric density of states is not justified.

I am very grateful to A. V. Balatsky 
and A. Chubukov for helpful discussions.
This work was supported by the 
National Science Foundation through Grant DMR-92-14739
and through the Materials Research Science 
and Engineering Center Program,
Grant No. DMR-96-32527. 

\section{Appendix}
In this appendix I give details of the more lengthy integrations.
Eq.\ \ref{eq:app1} is
\begin{equation}
{\cal G}(i \omega, r \hat{x}) =
- \frac{N_0}{4 \pi} \int d \phi 
\int d s \int_{-1}^1 e^{i u x}
\frac{i \Omega +  s \tau_3 + \cos(\phi)(1 - x^2)^{1/2} \tau_1}
{\Omega^2 + s^2 + \cos^2(\phi) (1 - x^2)} dx.
\label{eq:app1}
\end{equation}

The last term in the numerator gives zero on
integration over $\phi$.
We again write the x integral
as
\begin{eqnarray}
I & = & \int_{-1}^1 dx e^{i u x}
\frac{i \Omega +  s \tau_3 }
{\Omega^2 + s^2 + \cos^2(\phi) (1 - x^2)} \nonumber \\
& = & \oint dz~ e^{i u z}
\frac{i \Omega +   s \tau_3}
{\Omega^2 + s^2 + \cos^2(\phi) - z^2 \cos^2(\phi)}  \nonumber \\
& & - i \int_0^{\infty} dy~ e^{i u (1 + i y)}
\frac{i \Omega + s \tau_3}
{\Omega^2 + s^2 + \cos^2(\phi) - (1 + i y)^2 \cos^2(\phi)}  \nonumber \\
& & + i \int_0^{\infty} dy~ e^{i u (-1 + i y)}
\frac{i \Omega + s \tau_3}
{\Omega^2 + s^2 + \cos^2(\phi) - (-1 + i y)^2 \cos^2(\phi)}.
\end{eqnarray}
The poles lie on the real axis outside the contour, so 
the contour integral is zero.  Thus we obtain a very simple
result:
\begin{equation}
I = \frac{-i(i \Omega + s \tau_3)}
{\Omega^2 + s^2} \frac{e^{iu} - e^{-iu}}{u},
\end{equation}
again because $ u >> 1$.  Then
\begin{eqnarray}
{\cal G}(i \omega, r \hat{x}) & = &
 \frac{i N_0}{4 \pi k_F r} \int d \phi 
\int d s 
\frac{(i \Omega +  s \tau_3) ds}
{\Omega^2 + s^2} (e^{i (k_F + s \Delta_0/v_F) r} - 
e^{- i (k_F + s \Delta_0/v_F) r})\nonumber \\
  & = &  \frac{- \pi N_0}{2 k_F r}
        \left[ e^{i k_F r} (sgn~(\omega) \tau_0 + \tau_3) -  e^{- i k_F r} 
(sgn~(\omega) \tau_0  - \tau_3) \right].
\end{eqnarray}
This is Eq.\ \ref{eq:app2}.

Let us now carry out the integration for a polar
gap in the general case.
The Green's function is given by Eq.\ \ref{eq:app3}:
\begin{equation}
{\cal G}(r (\sin \psi \hat{x} + \cos \psi \hat{z}), i \omega)
= - \frac{N_0}{4 \pi}
\int_0^{2 \pi} d \phi \int d \epsilon \int_{-1}^1 dx
e^{iux} \frac{i \omega + \epsilon \tau_3 + \Delta_0 \tau_1
(ax - b \sqrt{1 - x^2})}{\omega^2 + \epsilon^2 + \Delta_0^2
(ax - b \sqrt{1 - x^2})^2},
\end{equation}
where $a \equiv \cos \psi, b = \cos \phi \sin \psi$,
and $u \equiv (k_F + \epsilon /v_F)r$.
The coordinates for the integration have been rotated so that
the polar axis defined by $x = \pm 1$ is along $\vec{r}$.
This is  rewritten as:
\begin{equation}
{\cal G}(r (\sin \psi \hat{x} + \cos \psi \hat{z}, i \omega)
= - \frac{N_0}{4 \pi \Delta_0}
\int_0^{2 \pi} d \phi \int ds I(\Omega,a,b),
\end{equation}
where
\begin{equation}
I(\Omega,a,b) = \int_{-1}^1 dx
e^{iux} \frac{i \Omega + s \tau_3 + \tau_1
(ax - b \sqrt{1 - x^2})}{\Omega^2 + s^2 + (ax - b \sqrt{1 - x^2})^2},
\end{equation}
which may again be evaluated using the same contour
as above.
This procedure yields
\begin{eqnarray}
I(\Omega,a,b)& = & \oint dz
e^{iuz} \frac{i \Omega + s \tau_3 + \tau_1
(az - b \sqrt{1 - z^2})}{\Omega^2 + s^2 + [az - b \sqrt{1 - z^2}]^2} \nonumber \\
  &  & - i  \int_{0}^{\infty} dy                           
e^{iu(1+iy)} \frac{i \Omega + s \tau_3 + \tau_1
(a(1+iy) - b \sqrt{1 - (1+iy)^2})}{\Omega^2 + s^2 + 
[a(1+iy) - b \sqrt{1 - (1+iy)^2}]^2}
 \nonumber \\
  &  &  i  \int_{0}^{\infty} dy 
e^{iu(-1+iy)} \frac{i \Omega + s \tau_3 + \tau_1
(a(-1+iy) - b \sqrt{1 - (-1+iy)^2})}{\Omega^2 + s^2 + 
[a(-1+iy) - b \sqrt{1 - (-1+iy)^2}]^2}.
\end{eqnarray}  
The poles lie at
points determined by the equation
\begin{equation}
a x - b \sqrt{1 - x^2} = \pm i (\Omega^2 + s^2)^{1/2}.
\end{equation}
There are four such roots:
\begin{equation}
x_{\pm,\pm} = \pm \frac{b (a^2 + b^2 +s^2 + \Omega^2)^{1/2}}
{(a^2 + b^2)}
\pm \frac{i a (\Omega^2 + s^2)^{1/2}}{a^2 + b^2}.
\end{equation}
Only the two with positive
imaginary parts ($x_{++}$ and $x_{-+}$) may lie in the contour,
and then only if $|Re~x| < 1$.  This is the case
if $|s| < (a^4/b^2 + a^2 - \Omega^2)^{1/2} \equiv s_0$.
The cuts
produced by the square roots can be chosen to be
along $(-\infty,-1)$ and $(1, \infty)$.  They lie outside the
contour.  
Performing the integrations leads to
\begin{eqnarray}
I(\Omega,a,b)  & = & \Theta(s_0 -|s|)
\pi e^{i u x_{++}} \frac{i \Omega + s \tau_3 + 
i \sqrt{\Omega^2 + s^2} \tau_1}
{ \sqrt{\Omega^2 + s^2}} \nonumber \\
  &  & + \Theta(s_0 -|s|) 
  \pi e^{i u x_{-+}} \frac{i \Omega + s \tau_3 + 
 i \sqrt{\Omega^2 + s^2} \tau_1}
{ \sqrt{\Omega^2 + s^2}}\nonumber \\
  &  & -i \frac{e^{i k_F r}}{k_F r}
  \frac{i \Omega + s \tau_3 + a \tau_1} {\Omega^2 + s^2 + a^2}
  e^{i s \Delta_0 r/v_F} \nonumber \\
  && +i \frac{e^{-i k_F r}}{k_F r}
  \frac{i \Omega + s \tau_3 - a \tau_1} {\Omega^2 + s^2 + a^2}
  e^{-i s \Delta_0 r/v_F}.
\end{eqnarray}    
This must next be integrated over $s$.  The first two integrals
can be performed by the stationary phase approximation,
taking $u$ as a large parameter and expanding the 
argument of the exponential about its maximum. 
These expansions are valid if $k_F r \sin \psi >>1$ 
and $k_F r \cos \psi >>1$.  
Thus, the result is not valid near the equatior or the poles.
Fortunately, we already
have results in these regions.  
The second                    
two integrals are standard contour integrations.
This leads to the expression
\begin{eqnarray}
\int_{-\infty}^{\infty} I(\Omega,a,b)  & = & 
2 \pi i (sgn~(\omega) + \tau_1)
[\frac{2 \pi |\Omega| (a^2 + b^2)}{au}]^{1/2}
\exp\left[\frac{iub \sqrt{a^2+b^2+\Omega^2}-au |\Omega|}{a^2+b^2}\right] \nonumber \\
& & + 2 \pi i (sgn~(\omega) + \tau_1)
[\frac{2 \pi |\Omega| (a^2 + b^2)}{au}]^{1/2}
\exp\left[\frac{-iub \sqrt{a^2+b^2+\Omega^2}-au |\Omega|}{a^2+b^2}\right] \nonumber \\
&& + \frac{\pi e^{i k_F r}}{k_F r}
[\frac{\Omega}{\sqrt{\Omega^2 + a^2}} 
- i \frac{a \tau_1}{\sqrt{\Omega^2 + a^2}} + \tau_3]
e^{-\sqrt{\Omega^2 + a^2} \Delta_0 r/v_F} \nonumber \\
&& - \frac{\pi e^{-i k_F r}}{k_F r}
[\frac{\Omega}{\sqrt{\Omega^2 + a^2}} 
+ i \frac{a \tau_1}{\sqrt{\Omega^2 + a^2}} - \tau_3]
e^{-\sqrt{\Omega^2 + a^2} \Delta_0 r/v_F}.
\end{eqnarray}    
The final integration over the azimuthal angle
is also simplified by the fact that $u>>1$,
and the same method may be used.
\newpage
\begin{eqnarray}
{\cal G}(r (\sin \psi \hat{x} + \cos \psi \hat{z}, i \omega)
& = & - \frac{N_0 \Delta_0}{4 \pi}
\int_0^{2 \pi} d \phi \int ds I(\Omega,a,b)\nonumber \\
  & = &   - \frac{i N_0 \Delta_0}{2}
  (sgn~(\Omega) + \tau_1) \int_0^{2 \pi}
  [\frac{2 \pi |\Omega| (\cos^2 \psi + \sin^2 \psi \cos^2 \phi)}
  {u \cos \psi}]^{1/2} \times 
  \nonumber \\
  &  &  \exp[\frac{i u \sin \psi \cos \phi -u |\Omega| \cos \psi}
           {\cos^2 \psi + \sin^2 \psi \cos^2 \phi}] d \phi \nonumber \\
  &  &   - \frac{i N_0 \Delta_0}{2}
  (sgn~(\Omega) + \tau_1) \int_0^{2 \pi}
  [\frac{2 \pi |\Omega| (\cos^2 \psi + \sin^2 \psi \cos^2 \phi)}
  {u \cos \psi}]^{1/2} \times 
  \nonumber \\
  &  &  \exp[\frac{- i u \sin \psi \cos \phi -u |\Omega| \cos \psi}
           {\cos^2 \psi + \sin^2 \psi \cos^2 \phi}] d \phi \nonumber \\
  &  &  + \frac{-i N_0 \pi e^{i k_F r}}{2 k_F r}
        [\frac{\omega}{\sqrt{\omega^2 + \Delta_0^2 \cos^2 \psi}} 
          - i \frac{\cos \psi \tau_1}
            {\sqrt{\omega^2 + \Delta_0^2 \cos^2 \psi}} 
              + \tau_3] \times \nonumber \\
  & &             e^{-\sqrt{\omega^2 + \Delta_0^2 \cos^2 \psi} r/v_F} \nonumber \\
  &  &  + \frac{i N_0 \pi e^{i k_F r}}{2  k_F r}
        [\frac{\omega}{\sqrt{\omega^2 + \Delta_0^2 \cos^2 \psi}} 
          + i \frac{\cos \psi \tau_1}{\sqrt{\omega^2 + \cos^2 \psi}} 
              - \tau_3] \times \nonumber \\
  & &      e^{-\sqrt{\omega^2 + \Delta_0^2 \cos^2 \psi} r/v_F}.
\end{eqnarray}  
The integrations then give
\begin{eqnarray}
{\cal G}(r (\sin \psi \hat{x} + \cos \psi \hat{z}, i \omega)
 & = &   - \frac{i \pi N_0}{2 k_F r}
  (sgn~(\omega) + \tau_1) 
    [\frac{2  |\omega|}
    { \Delta_0 |\cos \psi|^2 |\sin \psi|^2}]^{1/2} \times   \nonumber \\
  &  &  e^{i k_F r \sin \psi  
  - k_F r |\omega| |\cos \psi|/ \Delta_0} \nonumber \\
 & & - \frac{i \pi N_0}{2 k_F r}
  (sgn~(\omega) + \tau_1) 
    [\frac{2  |\omega|}
    { \Delta_0 |\cos \psi|^2 |\sin \psi|^2}]^{1/2} \times   \nonumber \\
  &  &  e^{- i k_F r \sin \psi  
  - k_F r |\omega| |\cos \psi|/ \Delta_0} \nonumber \\
  &  &  + \frac{-i N_0 \pi e^{i k_F r}}{2 k_F r}
        [\frac{\omega}{\sqrt{\omega^2 + \Delta_0^2 \cos^2 \psi}} 
          - i \frac{\cos \psi \tau_1}
          {\sqrt{\omega^2 + \Delta_0^2 \cos^2 \psi}} + \tau_3] \times \nonumber \\
  &  &  e^{-\sqrt{\omega^2 + \Delta_0^2 \cos^2 \psi} r/v_F} \nonumber \\
  &  &  + \frac{i N_0 \pi e^{i k_F r}}{2 k_F r}
        [\frac{\omega}{\sqrt{\omega^2 + \Delta_0^2 \cos^2 \psi}} 
          + i \frac{\cos \psi \tau_1}
          {\sqrt{\omega^2 + \Delta_0^2 \cos^2 \psi}} 
              - \tau_3] \times \nonumber \\
  &  &  e^{-\sqrt{\omega^2 + \Delta_0^2 \cos^2 \psi} r/v_F}.
\end{eqnarray}  
This leads immediately to Eq.\ \ref{eq:ph3}.

The Green's function for motion away from the impurity
along the x-axis in
two dimensions is given by Eq.\ \ref{eq:2dgx1}:
\begin{equation}
{\cal G}(i \omega, r \hat{x}) = - \frac{2 N_0}{\pi}
\int d \epsilon \int_{-1}^1 \frac{dv}{\sqrt{1-v^2}}
\frac{i \omega + \epsilon \tau_3 + \Delta_0 \tau_1 (2 v^2 -1)}
{\omega^2 + \epsilon^2 + \Delta_0^2 (2 v^2 -1)^2} e^{iuv}.
\label{eq:app5}
\end{equation}
Here $v = \cos(\phi)$, $ \Delta = \Delta_0 (2v^2-1)$ and
$u = (k_F + \epsilon/v_F) r$.  
We may also write
\begin{equation}
{\cal G}(i \omega, r \hat{x}) = - \frac{2 N_0}{\pi}
\int d s I(\Omega,s),
\end{equation}
where
\begin{equation}
I(\Omega,s) \equiv  \int_{-1}^1 \frac{dv}{\sqrt{1-v^2}}
\frac{i \Omega + s \tau_3 + \tau_1 (2 v^2 -1)}
{t^2 + (2 v^2 -1)^2} e^{iuv},
\end{equation}
and $\Omega = \omega/\Delta_0$, $s = \epsilon/\Delta_0$,
and $t = \sqrt{\Omega^2 + s^2}$.  As usual, we split the 
integral into three parts:
\begin{equation}
I(\Omega,s) = I_1(\Omega,s) + I_2(\Omega,s)+ I_3(\Omega,s), 
\end{equation}
with
\begin{equation}
I_1(\Omega,s) = \oint \frac{dz}{\sqrt{1-z^2}}
\frac{i \Omega + s \tau_3 + \tau_1 (2 z^2 -1)}
{t^2 + (2 z^2 -1)^2} e^{iuz},
\end{equation}
\begin{equation}
I_2(\Omega,s) = -i \int_0^{\infty} \frac{dy}{\sqrt{1-(1+iy)^2}}
\frac{i \Omega + s \tau_3 + \tau_1 [2 (1+iy)^2 -1]}
{t^2 + [2 (1+iy)^2 -1]^2} e^{iu(1+iy)},
\end{equation}
and
\begin{equation}
I_3(\Omega,s) = i \int_0^{\infty} \frac{dy}{\sqrt{1-(-1+iy)^2}}
\frac{i \Omega + s \tau_3 + \tau_1 [2 (-1+iy)^2 -1]}
{t^2 + [2 (-1+iy)^2 -1]^2} e^{iu(-1+iy)}.
\end{equation}
Turning first to $I_1$, we locate the poles of the integrand at
\begin{equation}
t^2 + (2 z^2 -1 )^2 = 0
\end{equation}
or
\begin{equation}
z^2 = \frac{1}{2} ( 1 \pm it).
\end{equation}
There are four poles, only two of which have positive
imaginary parts and thus may lie in the contour.
Let us call them $z_1$ and $z_2$.  They satisfy
\begin{equation}
|z_{1}| = |z_{2}| = |z| 
\frac{1}{\sqrt{2}} (1+ t^2)^{1/4}
\end{equation}
and $ z_{1}  = - |z| e^{- i \theta}$, 
$ z_{2}  = |z| e^{i \theta}$, with $\theta = \frac{1}{2} \tan^{-1}(t)$.
If the real parts of these quantities are less than unity in absolute
magnitude, then they lie in the contour.  A bit of trigonometry shows that this occurs
only if $t < 2 \sqrt{2}$.  This condition will normally be fulfilled
in our problem.  The other two roots lie at $ z_3 = - |z| e^{ i \theta}$
and $ z_4 = |z| e^{- i \theta}$.

The contour integral is now written as :
\begin{equation}
I_1(\Omega,s) = \frac{1}{4} \oint \frac{dz}{\sqrt{1-z^2}}
\frac{i \Omega + s \tau_3 + \tau_1 (2 z^2 -1)}
{(z-z_1)(z-z_2)(z-z_3)(z-z_4)} e^{iuz},
\end{equation}
and the residue theorem gives
\begin{eqnarray}
I_1(\Omega,s) & = & \frac{1}{4} 2 \pi i \Theta(\sqrt{8}- t) \times \nonumber \\
&& \left[ 
\frac{i \Omega + s \tau_3 + \tau_1 (2 z_1^2 -1)}
{\sqrt{1-z_1^2}(z_1-z_2)(z_1-z_3)(z_1-z_4)} e^{iuz_1}
+ \frac{i \Omega + s \tau_3 + \tau_1 (2 z_2^2 -1)}
{\sqrt{1-z_2^2}(z_2-z_1)(z_2-z_3)(z_2-z_4)} e^{iuz_2} \right].
\end{eqnarray}
Using
\begin{eqnarray}
\sqrt{1-z_1^2}  & = 
& \frac{1}{\sqrt{2}} (1+t^2)^{1/4} e^{i \theta} 
= |z| e^{i \theta},\nonumber \\
\sqrt{1-z_2^2}  & = & \frac{1}{\sqrt{2}} (1+t^2)^{1/4} e^{- i \theta}
= |z| e^{- i \theta},\nonumber \\
2z_1^2 - 1 & = & - it,\nonumber \\
2z_2^2 - 1 & = &  it,
\end{eqnarray} 
this may be rewritten as
\begin{eqnarray}
I_1(\Omega,s) & = &  \frac{ i \pi}{2} \Theta(\sqrt{8}- t) |z|^{-4} \times \nonumber \\ 
&& [\frac{(i \Omega + s \tau_3 - i t \tau_1) e^{iuz_1}}
{e^{i \theta} (- e^{- i \theta} - e^{i \theta})
(- e^{- i \theta} + e^{i \theta})( - e^{-i \theta} - e^{-i \theta})} \nonumber \\
& & + \frac{(i \Omega + s \tau_3 + i t \tau_1)e^{iuz_2}}
{e^{-i \theta} (e^{ i \theta} + e^{- i \theta})
(e^{ i \theta} + e^{i \theta})( e^{i \theta} - e^{-i \theta})}] \nonumber \\
& = & \frac{ i \pi}{2} \Theta(\sqrt{8}- t) (\frac{4}{1+t^2}) \times \nonumber \\
& & [ \frac{(i \Omega + s \tau_3 - i t \tau_1) e^{iuz_1}}
{(- 2 \cos \theta )
(2 i \sin \theta)( - 2 )} + \frac{(i \Omega + s \tau_3 + i t \tau_1)e^{iuz_2}}
{ (2 \cos\theta)
(2)( 2 i \sin \theta )}].
\end{eqnarray}
Since $ \frac{1}{2} \sin \theta \cos \theta \sin 2 \theta = 
\sin \tan^{-1} t = t/ \sqrt{1+t^2}$, 
this becomes
\begin{eqnarray}
I_1(\Omega,s) & = &  
\frac{ \pi}{2} \Theta(\sqrt{8}- t) (\frac{1}{t \sqrt{1+t^2}}) \times \nonumber \\
& & [ (i \Omega + s \tau_3 - i t \tau_1) e^{iuz_1}
 + (i \Omega + s \tau_3 + i t \tau_1)e^{iuz_2} ].
\end{eqnarray}
Integrating this:
\begin{eqnarray}
\int ds I_1(\Omega,s) & = & 
\frac{ \pi}{2} \int_{-\infty}^{\infty} ds
\Theta(\sqrt{8}- \sqrt{\Omega^2+s^2}) (\frac{1}{\sqrt{\Omega^2+s^2} 
\sqrt{1 + \Omega^2+s^2}}) \times  \nonumber \\
& & \left[ (i \Omega + s \tau_3 - i \sqrt{\Omega^2+s^2} \tau_1) \right.
\exp\left(\frac{-iu}{2}\sqrt{\sqrt{1+\Omega^2+s^2}+1} \right) \times \nonumber \\
& & \exp\left(\frac{-u}{2}\sqrt{\sqrt{1+\Omega^2+s^2}-1}\right) \nonumber \\
& &  + (i \Omega + s \tau_3 + i \sqrt{\Omega^2+s^2} \tau_1)
\exp\left(\frac{iu}{2}\sqrt{\sqrt{1+\Omega^2+s^2}+1}\right) \times \nonumber \\
& & \left. \exp\left((\frac{-u}{2}\sqrt{\sqrt{1+\Omega^2+s^2}-1}\right)\right].
\end{eqnarray}
Since $ u >> 1$, we expand the argument of the exponential around its
maximum at $s=0$:
\begin{eqnarray}
\exp(\frac{-u}{2}\sqrt{\sqrt{1+\Omega^2+s^2}-1}) 
& \approx & 
\exp(\frac{-u}{2}\sqrt{\sqrt{1+\Omega^2}-1}) \times \nonumber \\
& & \exp[\frac{-us^2}{8}(\sqrt{1+\Omega^2}-1)^{-1/2}(1+\Omega^2)^{-1/2}],
\end{eqnarray}
and evaluate the rest of the integrand at $s = 0 $,
so that $ t \rightarrow \Omega$.  This yields
\begin{eqnarray}
\int ds I_1(\Omega,s) & = & 
\frac{ \pi}{2}  (\frac{(\sqrt{1+\Omega^2} -1 )^{1/4} }
{(1+\Omega^2)^{1/4}}) (\frac{8 \pi }{k_F r})^{1/2} 
\exp(\frac{-u}{2}\sqrt{\sqrt{1+\Omega^2}-1}) 
 \times  \nonumber \\
& & [ (i ~sgn~\Omega - i \tau_1) 
\exp(\frac{-iu}{2}\sqrt{\sqrt{1+\Omega^2}+1}) \nonumber \\
& &  + (i sqn~\Omega + i \tau_1)
\exp(\frac{iu}{2}\sqrt{\sqrt{1+\Omega^2}+1})].
\label{eq:app6}
\end{eqnarray}

The other integrals are simpler:
\begin{eqnarray}
I_2(\Omega,s) & = & -i e^{iu} \int \frac{dy}{\sqrt{y^2 - 2 i y}}
\frac{i \Omega + s \tau_3 + \tau_1 [2 (1+iy)^2 -1]}
{t^2 + [2 (1+iy)^2 -1]^2} e^{-uy} \nonumber \\
& \approx & 
-i (-2i)^{-1/2} e^{iu} \frac{i \Omega + s \tau_3 + \tau_1}{t^2+1}
\int_0^{\infty} \frac{dy}{\sqrt{y}}e^{-uy} \nonumber \\
& = & 
\frac{1-i}{2} \sqrt{\frac{\pi}{k_Fr}} e^{ik_Fr} e^{is\Delta_0 r/v_F}
\frac{i \Omega + s \tau_3 + \tau_1}{s^2+\Omega^2+1},
\end{eqnarray}
and similarly
\begin{equation}
I_3(\Omega,s)
\frac{1+i}{2} \sqrt{\frac{\pi}{k_Fr}} e^{-ik_Fr} e^{-is\Delta_0 r/v_F}
\frac{i \Omega + s \tau_3 + \tau_1}{s^2+\Omega^2+1}.
\end{equation}
These expression are easily integrated over energy:
\begin{equation}
\int_{-\infty}^{\infty} ds I_2(\Omega,s) =
\frac{(1-i)\pi}{2} \sqrt{\frac{\pi}{k_Fr}} e^{ik_Fr} 
e^{-\sqrt{1+\Omega^2}\Delta_0 r/v_F}
\frac{i \Omega + i \sqrt{1+\Omega^2} \tau_3 + \tau_1}{\sqrt{1+\Omega^2}},
\label{eq:app7}
\end{equation}
and
\begin{equation}
\int_{-\infty}^{\infty} ds I_3(\Omega,s) =
\frac{(1+i)\pi}{2} \sqrt{\frac{\pi}{k_Fr}} e^{-ik_Fr} 
e^{-\sqrt{1+\Omega^2}\Delta_0 r/v_F}
\frac{i \Omega - i \sqrt{1+\Omega^2} \tau_3 + \tau_1}{\sqrt{1+\Omega^2}}.
\label{eq:app8}
\end{equation}
\newpage
Combining Eqs.\ \ref{eq:app5}, \ref{eq:app6}, \ref{eq:app7},
and \ref{eq:app8},
we find
\begin{eqnarray}
{\cal G}(i \omega, r \hat{x}) & = & - \frac{2 N_0}{\pi} \times \nonumber \\
  &  & [ \frac{ \pi}{2}  (\frac{(\sqrt{1+\Omega^2} -1 )^{1/4} }
{(1+\Omega^2)^{1/4}}) (\frac{8 \pi }{k_F r})^{1/2} 
\exp(\frac{-u}{2}\sqrt{\sqrt{1+\Omega^2}-1}) 
 \times  \nonumber \\
& & [ (i ~sgn~\Omega - i \tau_1) 
\exp(\frac{-iu}{2}\sqrt{\sqrt{1+\Omega^2}+1}) \nonumber \\
& &  + (i sqn~\Omega + i \tau_1)
\exp(\frac{iu}{2}\sqrt{\sqrt{1+\Omega^2}+1})] \nonumber \\
  &  & + \frac{(1-i)\pi}{2} \sqrt{\frac{\pi}{k_Fr}} e^{ik_Fr} 
e^{-\sqrt{1+\Omega^2}\Delta_0 r/v_F}
\frac{i \Omega + i \sqrt{1+\Omega^2} \tau_3 + \tau_1}{\sqrt{1+\Omega^2}},\nonumber \\
& &  + \frac{(1-i)\pi}{2} \sqrt{\frac{\pi}{k_Fr}} e^{-ik_Fr} 
e^{-\sqrt{1+\Omega^2}\Delta_0 r/v_F}
\frac{i \Omega - i \sqrt{1+\Omega^2} \tau_3 + \tau_1}{\sqrt{1+\Omega^2}}.\nonumber \\
\end{eqnarray}
From this Eq.\ \ref{eq:giwrx} follows immediately.

The Green's function for the 
direction along the node is more easily computed.
We have, from Eq.\ \ref{eq:2dgx2}, that
\begin{equation}
{\cal G}(i \omega, r \frac{\hat{x} + \hat{y}}{\sqrt{2}}) = - \frac{2 N_0}{\pi}
\int d \epsilon \int_{-1}^1 \frac{dv}{\sqrt{1-v^2}}
\frac{i \omega + \epsilon \tau_3 + \Delta_0 \tau_1 v \sqrt{1-v^2}}
{\omega^2 + \epsilon^2 + 2 \Delta_0^2 v^2(1- v^2)} e^{iuv},
\label{eq:app9}
\end{equation}
where the part proportional to $\tau_1$
vanishes by symmetry.  
and with the usual breakup:
\begin{equation}
{\cal G}(i \omega, r \frac{\hat{x} + \hat{y}}{\sqrt{2}}) = - \frac{2 N_0}{\pi}
\int ds (I_1 + I_2 + I_3). 
\label{eq:gsum}
\end{equation}
Now
\begin{eqnarray}
I_1 & = & \oint \frac{dz}{\sqrt{1-z^2}}
\frac{i \Omega + s \tau_3}
{\Omega^2 + s^2 + 2 z^2(1- z^2)} e^{iuz} \nonumber \\
& = & -\frac{1}{4} \oint 
\frac{i \Omega + s \tau_3 + \tau_1 z \sqrt{1-z^2}}
{(z-z_1)(z-z_2)(z-z_3)(z-z_4)} e^{iuz},
\end{eqnarray}
where the roots are
\begin{eqnarray}
z_1  & = & \frac{i}{\sqrt{2}} (\sqrt{1+t^2}-1)^{1/2}  \nonumber \\
z_2  & =  & - z_1 \nonumber \\
z_3  & =  & - \frac{1}{\sqrt{2}} (\sqrt{1+t^2}+1)^{1/2}\nonumber \\
z_4  & =  & - z_3.
\end{eqnarray}
We shall also need:
\begin{equation}
\sqrt{1-z_1^2} = z_4
\end{equation}
Only $z_1$ lies in the contour.
Evaluating the integral:
\begin{eqnarray}
I_1  & =  &  - \frac{2 \pi i}{4} (1-z_1^2)^{-1/2}
\frac{i \Omega + s \tau_3}
{(z_1-z_2)(z_1-z_3)(z_1-z_4)} e^{iuz_1}\nonumber \\
  & =  & \frac{- \pi i}{2} 
\frac{i \Omega + s \tau_3}
{z_4 (2 z_1)(z_1^2-z_3^2)}e^{iuz_1}\nonumber \\
  & =  & \frac{\pi
(i \Omega + s \tau_3}
{|t| \sqrt{1+t^2}} \exp[\frac{-u}{\sqrt{2}} (\sqrt{1+t^2}-1)^{1/2} ]
\label{eq:int1}
\end{eqnarray}
Integrating over energy:
\begin{eqnarray}
\int ds I_1  & =  & \pi \int^{\infty}_{-\infty} ds
\frac{i \Omega + s \tau_3}
{\sqrt{\Omega^2 + s^2} \sqrt{1+\Omega^2 + s^2}}
\exp[\frac{-u}{\sqrt{2}} (\sqrt{1+\Omega^2+s^2}-1)^{1/2} ]\nonumber \\
& = & \pi \frac{i \Omega}
{|\Omega| \sqrt{1+\Omega^2 }}
\exp[\frac{-u}{\sqrt{2}} (\sqrt{1+\Omega^2}-1)^{1/2}] \times \nonumber \\
& & \int^{\infty}_{-\infty} ds
\exp[\frac{-us^2}{4 \sqrt{2}} (\sqrt{1+\Omega^2}-1)^{1/2} \sqrt{1+\Omega^2}] \nonumber \\
& = & (\frac{4 \pi^3 \sqrt{2}}{k_F r})^{1/2}
\frac{i ~sgn~\Omega}{(1+\Omega^2)^{1/4}}  (\sqrt{1+\Omega^2}-1)^{1/4}
e^{-k_F r  (\sqrt{1+\Omega^2}-1)^{1/2}/\sqrt{2}}.
\label{eq:app10}
\end{eqnarray}
The other integrals are simpler, as usual.
\begin{eqnarray}
I_2  & = & -i \int_{0}^{\infty} \frac{dy}{\sqrt{1-(1+iy)^2}}
\frac{i \Omega + s \tau_3 }
{\Omega^2 + s^2 + 2 (1+iy)^2(1- (1+iy)^2)} e^{iu(1+iy)}, \nonumber \\
  & \approx &  -i e^{iu} \int_{0}^{\infty} \frac{dy}{\sqrt{-2iy}} e^{-uy}
\frac{i \Omega + s \tau_3 }
{s^2+ \Omega^2} \nonumber \\
 & = & \frac{1-i}{2}~\frac{e^{ik_Fr}}{\sqrt{k_Fr}}~\frac{i \Omega + s \tau_3}
{s^2 + \Omega^2} e^{i k_F s \Delta_0 r/v_F},
\end{eqnarray}
and
\begin{eqnarray}
I_3  & = & i \int_{0}^{\infty} \frac{dy}{\sqrt{1-(-1+iy)^2}}
\frac{i \Omega + s \tau_3 }
{\Omega^2 + s^2 + 2 (-1+iy)^2(1- (-1+iy)^2)} e^{iu(-1+iy)}, \nonumber \\
  & \approx &  i e^{-iu} \int_{0}^{\infty} \frac{dy}{\sqrt{2iy}} e^{-uy}
\frac{i \Omega + s \tau_3 }
{s^2+ \Omega^2} \nonumber \\
 & = & \frac{1+i}{2}~\frac{e^{-ik_Fr}}{\sqrt{k_Fr}}~\frac{i \Omega + s \tau_3}
{s^2 + \Omega^2} e^{-i k_F s \Delta_0 r/v_F}.
\end{eqnarray}
The energy integrals are easily done:
\begin{equation}
\int ds I_2 = \frac{(1-i) \pi}{2}~\frac{e^{ik_Fr}}{\sqrt{k_Fr}}~(i ~sgn~\Omega + i \tau_3)
e^{- |\omega| r/v_F},
\label{eq:app11}
\end{equation}
and
\begin{equation}
\int ds I_3 = \frac{(1+i) \pi}{2}~\frac{e^{-ik_Fr}}{\sqrt{k_Fr}}~(i ~sgn~\omega - i \tau_3)
e^{- |\omega| r/v_F}.
\label{eq:app12}
\end{equation}
From Eqs.\ \ref{eq:gsum}, \ref{eq:app10}, 
\ref{eq:app11}, and \ref{eq:app12}, the Green's function is
\begin{eqnarray}
{\cal G}(i \omega, r \frac{\hat{x} 
+ \hat{y}}{\sqrt{2}}) &  = & - \frac{2 N_0}{\pi} \times \nonumber \\
  &  & [ (\frac{4 \pi^3 \sqrt{2}}{k_F r})^{1/2}
\frac{i ~sgn~\Omega}{(1+\Omega^2)^{1/4}}  (\sqrt{1+\Omega^2}-1)^{1/4}
e^{-k_F r  (\sqrt{1+\Omega^2}-1)^{1/2}/\sqrt{2}} \nonumber \\
  &  & + \frac{(1-i) \pi}{2}~\frac{e^{ik_Fr}}{\sqrt{k_Fr}}~(i ~sgn~\Omega + i \tau_3)
e^{- |\omega| r/v_F}\nonumber \\
  &  & + \frac{(1+i) \pi}{2}~\frac{e^{-ik_Fr}}{\sqrt{k_Fr}}~(i ~sgn~\omega - i \tau_3)
e^{- |\omega| r/v_F}].
\label{eq:app14}
\end{eqnarray}
From this, Eq.\ \ref{eq:grxy} follows immediately.

The general direction is given by the equation:
\begin{eqnarray}
{\cal G}(i \omega, r \cos(\frac{\pi}{4} - \psi)\hat{x} + 
r \sin(\frac{\pi}{4} - \psi) \hat{y}) &  
= & - \frac{ N_0}{\pi} \times \nonumber \\
& & \int d \epsilon \int_{-1}^1 \frac{dv}{\sqrt{1-v^2}}
\frac{i \omega + \epsilon \tau_3 + \Delta_0 \tau_1 v \sqrt{1-v^2}}
{\omega^2 + \epsilon^2 + 2 \Delta_0^2 v^2(1- v^2)} \times \nonumber \\
& & e^{iu(v \cos \psi + \sqrt{1-v^2} \sin \psi)} \nonumber \\
  &  &  + (\sin \psi \rightarrow - \sin \psi,~\tau_1 \rightarrow -\tau_1).\nonumber \\
& &  = \int ds (I_1 + I_2 + I_3) 
\end{eqnarray} 
The two integrals come from the regions $\phi = (0, \pi)$ and $\phi = (\pi, 2 \pi)$.
Then we find
\begin{eqnarray}
I_1 & = & \frac{\pi}{2}\frac{i \Omega + s \tau_3 + i |t| \tau_1}{|t| \sqrt{1+t^2}}
\exp[iu(z_1 \cos \psi + z_4 \sin \psi)] \nonumber \\
  & + &  \frac{\pi}{2}\frac{i \Omega + s \tau_3 - i |t| \tau_1}{|t| \sqrt{1+t^2}}
\exp[iu(z_1 \cos \psi - z_4 \sin \psi)], 
\end{eqnarray}
where the result is obtained by the same method as that in Eq.\ \ref{eq:int1}.
The energy integral is:
\begin{eqnarray}
\int ds I_1  & = &  
 (\frac{4 \pi^3 \sqrt{2}}{k_F r |\cos \psi|})^{1/2}
\frac{i ~sgn~\Omega + i \tau_1}{(1+\Omega^2)^{1/4}}  (\sqrt{1+\Omega^2}-1)^{1/4} \times \nonumber \\
& & e^{-k_F r  |\cos \psi| (\sqrt{1+\Omega^2}-1)^{1/2}/\sqrt{2}}
\exp[i k_F r \sin \psi (\sqrt{1+\Omega^2}-1)^{1/2}/\sqrt{2}] \nonumber \\
  & + &  (\frac{4 \pi^3 \sqrt{2}}{k_F r |\cos \psi|})^{1/2}
\frac{i ~sgn~\Omega - i \tau_1}{(1+\Omega^2)^{1/4}}  (\sqrt{1+\Omega^2}-1)^{1/4} \times \nonumber \\
& & e^{-k_F r  |\cos \psi| (\sqrt{1+\Omega^2}-1)^{1/2}/\sqrt{2}}
\exp[- i k_F r \sin \psi (\sqrt{1+\Omega^2}-1)^{1/2}/\sqrt{2}].
\end{eqnarray}
Similarly, from Eqs.\ \ref{eq:app11} and \ref{eq:app12}, we have
\begin{equation}
\int ds I_2 = \frac{(1-i) \pi}{2}~\frac{e^{ik_Fr \cos \psi}}
{\sqrt{k_Fr|\cos \psi|}}~(i ~sgn~\Omega + i \tau_3)
e^{- |\omega \cos \psi| r/v_F},
\end{equation}
and
\begin{equation}
\int ds I_3 = \frac{(1+i) \pi}{2}~\frac{e^{-ik_Fr \cos \psi}}
{\sqrt{k_Fr |\cos \psi|}}~(i ~sgn~\omega - i \tau_3)
e^{- |\omega \cos \psi| r/v_F}.
\end{equation}
Finally, let us define
\begin{equation}
{\cal G}_-(i \omega, \psi)
= {\cal G}(i \omega, r \cos(\frac{\pi}{4} - \psi)\hat{x} + 
\sin(\frac{\pi}{4} - \psi) \hat{y}). 
\end{equation}

Collecting results, we have
\begin{eqnarray}
{\cal G}_-(i \omega, \psi)
&  = & - \frac{2 N_0}{\pi} \times \nonumber \\
& & \left\{
 (\frac{4 \pi^3 \sqrt{2}}{k_F r |\cos \psi|})^{1/2} \right.
\frac{i ~sgn~\Omega + i \tau_1}{(1+\Omega^2)^{1/4}}  
(\sqrt{1+\Omega^2}-1)^{1/4} \times \nonumber \\
& & e^{-k_F r  |\cos \psi| (\sqrt{1+\Omega^2}-1)^{1/2}/\sqrt{2}}
\exp\left[i k_F r \sin \psi (\sqrt{1+\Omega^2}-1)^{1/2}/\sqrt{2}\right] 
\nonumber \\
& + & (\frac{4 \pi^3 \sqrt{2}}{k_F r |\cos \psi|})^{1/2}
\frac{i ~sgn~\Omega - i \tau_1}{(1+\Omega^2)^{1/4}}  
(\sqrt{1+\Omega^2}-1)^{1/4} \times \nonumber \\
& & e^{-k_F r  |\cos \psi| (\sqrt{1+\Omega^2}-1)^{1/2}/\sqrt{2}}
\exp \left[- i k_F r \sin \psi (\sqrt{1+\Omega^2}-1)^{1/2}/
\sqrt{2}\right]\nonumber \\
& + & 
\frac{(1-i) \pi}{2}~\frac{e^{ik_Fr \cos \psi}}
{\sqrt{k_Fr|\cos \psi|}}~(i ~sgn~\Omega + i \tau_3)
e^{- |\omega \cos \psi| r/v_F} \nonumber \\
& + &
\left. \frac{(1+i) \pi}{2}~\frac{e^{-ik_Fr \cos \psi}}
{\sqrt{k_Fr |\cos \psi|}}~(i ~sgn~\omega - i \tau_3)
e^{- |\omega \cos \psi| r/v_F}] \right\}.
\end{eqnarray} 
This is Eq.\ \ref{eq:grpsi}.

\begin{figure}
\caption[]{Density of states for a model semiconductor with a band
structure which is symmetric about the gap midpoint.  This system contains 
a single impurity which is a unitary scatterer.  This results in
a midgap state.}
\label{fig:symm}
\end{figure}
\begin{figure}
\caption[]{Density of states for a  model semiconductor with a band
structure which is not symmetric about the gap midpoint.  
This system contains 
a single impurity which is a unitary scatterer.  This results in
a bound state which is not at midgap.  This is the generic case. }
\label{fig:asymm1}
\end{figure}
\begin{figure}
\caption[]{Density of states for an artifical semiconductor
with nearest-neighbor tight-binding dispersion and a gap.  This
density of states has an asymmetry factor of 0.25 t / unit cell.
The asymmetry factor is defined in the text.}
\label{fig:tbgap}
\end{figure}
\begin{figure}
\caption[]{Density of states for a semiconductor with many impurities
using the usual impurity averaging procedure.
The impurity density is above the critical value, so the gap has filled in.
The even density of states is for a semiconductor with a band structure
which is symmetric about the gap midpoint.   The parameters, referring to Figure
1, are $\epsilon_1/\Delta = 5$ and $n_{imp}V^2 N_0/\Delta = 0.8.$
The asymmetric denstiy of states is for an asymmetric density of states.
The parameters, referring to Fig. 2, are $\epsilon_{\ell}/\Delta = 5$,
$\epsilon_{u}/\Delta = 8$, and $n_{imp}V^2 N_0/\Delta = 0.8.$}
\label{fig:num}
\end{figure}
\begin{figure}
\caption[]{The density of states for a model semiconductor
with many impurities.  The calculation combines the results of
the T-matrix and the impurity averaging.  The concentration is subcritical,
so the gap is not completely closed.  The bound states are shown in black.  
Note that the total number of bound states
is proportional to the number of impurities, whereas the number of states in
a fixed energy range around the chemical potential is proportional to the 
total number of orbitals. }
\label{fig:sem1}
\end{figure}
\begin{figure}
\caption[]{The density of states for a model semiconductor
with many impurities.  The calculation combines the results of
the T-matrix and the impurity averaging.  The concentration is supercritical,
so the gap is closed.  The bound states, shown in black, become resonances.  
Note that the total number of states in the resonance peak
is proportional to the number of impurities, whereas the number of states in
a fixed energy range around the chemical potential is proportional to the 
total number of orbitals. }
\label{fig:sem2}
\end{figure}
\begin{figure}
\caption[]{Contour for the angular integrals involved in calculating the 
impurity wavefunction in real space. }
\label{fig:int}
\end{figure}\begin{figure}
\caption[]{Diagram (a), which contains no crossed impurity lines,
is counted in the usual calculations of transport properties
in unconventional superconductors.  Diagram (b), with crossed lines,
is usually neglected. }
\label{fig:non1}
\end{figure}
\begin{figure}
\caption[]{The diagrams of the previous figure in momentum space
in the normal state.  The justification for the neglect of
(b) is that one of the momenta is off the Fermi surface unless
an additional constraint is applied. }
\label{fig:non2}
\end{figure}
\begin{figure}
\caption[]{The same diagrams in momentum space, but now
in the superconducting state at low temperatures.  
All momenta must be near the nodes, which are situated on the diagonals.
The justification for the neglect of
(b) no longer applies.}
\label{fig:non3}
\end{figure}
\begin{figure}
\caption[]{The density of states for a d-wave superconductor
with many impurities.  The calculation combines the results of
the T-matrix and the impurity averaging.  The broadening of the resonance is
shown.  Note that the total number of states in the resonance peak
is proportional to the number of impurities, whereas the number of states in
a fixed energy range around the chemical potential is proportional to the 
total number of orbitals. }
\label{fig:smany}
\end{figure}

\end{document}